# Theoretical and Practical Limits of Kolmogorov-Zurbenko Periodograms with Dynamic Smoothing in Estimating Signal Frequencies

**Barry Loneck*, Igor Zurbenko, and Edward Valachovic**

**Department of Epidemiology and Biostatistics**
**School of Public Health**
**The University at Albany**
**State University of New York**

## ABSTRACT

This investigation establishes the theoretical and practical limits of Kolmogorov-Zurbenko periodograms with dynamic smoothing in their estimation of signal frequencies in terms of their sensitivity, accuracy, resolution, and robustness. While the DiRienzo-Zurbenko algorithm performs dynamic smoothing based on local variation in a periodogram, the Neagu-Zurbenko algorithm performs dynamic smoothing based on local departure from linearity in a periodogram. This article begins with a summary of the statistical foundations for both the DiRienzo-Zurbenko algorithm and the Neagu-Zurbenko algorithm, followed by instructions for accessing and utilizing these approaches within the *R* statistical program platform. Brief definitions, importance, statistical bases, theoretical and practical limits, and demonstrations are provided for their sensitivity, accuracy, resolution, and robustness in estimating signal frequencies. Next using a simulated time series in which two signals close in frequency are embedded in a significant level of random noise, the predictive power of these approaches are compared to the autoregressive integral moving average (ARIMA) approach, with support again garnered for their being robust when data is missing. Throughout, the article contrasts the limits of Kolmogorov-Zurbenko periodograms with dynamic smoothing to those of log-periodograms with static smoothing, while also comparing the performance of the DiRienzo-Zurbenko algorithm to that of the Neagu-Zurbenko algorithm. It concludes by delineating next steps to establish the precision with which Kolmogorov-Zurbenko periodograms with dynamic smoothing estimate signal strength.

## KEYWORDS

---

* Corresponding Author Email: bloneck@albany.edu

# Theoretical and Practical Limits of Kolmogorov-Zurbenko Periodograms with Dynamic Smoothing in Estimating Signal Frequencies

In sharp contrast to traditional periodograms that use static smoothing with a fixed window width, Kolmogorov-Zurbenko periodograms with dynamic smoothing use a window that varies in width across the frequency spectrum. At present, there are two principal dynamic smoothing methods: the DiRienzo-Zurbenko algorithm and the Neagu-Zurbenko algorithm. This investigation establishes their theoretical and practical limits in estimating signal frequencies in terms of sensitivity (i.e., ability to detect weak signals), accuracy (i.e., ability to correctly identify signal frequencies), resolution (i.e., ability to separate signals with close frequencies), and robustness (i.e., maintaining sensitivity, accuracy, and resolution when data is missing), then demonstrates their performance with respect to these limits using simulated datasets.

Simply put, a time series tracks the value of a given variable over time and time series analyses focus on how that variable changes with respect to time. Some time series analyses assume a linear model and its most common algorithm is the autoregressive integrated moving average (ARIMA). However, the linear model underlying these algorithms limits their predictive power when periodicity is present and trends in the data are better described as signals that are sinusoidal in nature. For a full discussion, see the texts of Koopmans (1995), Montgomery, Jennings, and Kulahci (2015), and Wei (2006).

Time series analyses based on sinusoidal models have two aspects. The first is the temporal dimension and it examines regularity in patterns with which the given variable changes over time. To extract these patterns, temporal analyses use band pass filters that typically utilize a moving average with a fixed width to extract the true signal from an otherwise noisy background of random fluctuations in the variable (Zurbenko, 1986). However, a better filter is the Kolmogorov-Zurbenko adaptive filter (Yang & Zurbenko, 2010; Zurbenko, 1991). While this filter also utilizes a moving average, it does so by varying the width of the filter's window such that it more accurately eliminates random noise and the actual variation in the variable over time (i.e., periodicity) can be more clearly seen.

The second aspect of time series analyses based on sinusoidal models is the spectral dimension and it searches for signals present in a given time series. To find these signals, spectral analyses utilize periodograms that are smoothed using a moving average with a fixed window width to identify specific signal frequencies present and their respective strength across the range of all possible frequencies. However, a better periodogram is the Kolmogorov-Zurbenko periodogram with dynamic smoothing, which includes both DiRienzo-Zurbenko algorithm smoothing (DiRienzo, 1996; DiRienzo & Zurbenko, 1999) and Neagu Zurbenko algorithm smoothing (Neagu & Zurbenko, 2003; Zurbenko, 2001). These periodograms, likewise, use a moving average but with window widths that are dynamic and vary across increments of the frequency spectrum. The DiRienzo-Zurbenko algorithm adjusts window width based on the proportion of total variance present such that it can precisely pinpoint the frequency of each signal in the time series. Alternatively, the Neagu-Zurbenko algorithm adjusts window width based on proportion of total departure from linearity present, leading to its ability to eliminate specious peaks and achieve better resolution. Consequently, the purpose of this investigation is to establish the theoretical and practical limits of Kolmogorov-Zurbenko periodograms with dynamic smoothing with respect to sensitivity, accuracy, resolution, and robustness in estimating signal frequencies.

This article begins with a definition of time series as a stochastic process, followed by a brief exploration of the statistical foundations of a standard spectral analysis along with its subsequent refinement with the Kolmogorov-Zurbenko periodogram in concert with the DiRienzo-Zurbenko algorithm. This is followed by a brief discussion of the statistical foundations for the Neagu-Zurbenko algorithm, including its key advantages over the DiRienzo-Zurbenko algorithm. Instructions for accessing and utilizing these algorithms are also provided. To establish their sensitivity, accuracy, resolution, and robustness in estimating signal frequencies, we provide brief definitions, importance, statistical bases, theoretical and practical limits, and demonstrations for each. Using a simulated time series dataset with relatively weak signals embedded in significant levels of noise, the article then presents a further demonstration of the sensitivity, accuracy, resolution, and robustness of the Kolmogorov-Zurbenko periodogram with dynamic smoothing, in addition to showcasing its predictive power in contrast to the ARIMA model. Throughout the article, we contrast dynamic smoothing with static smoothing, in theory, as well as compare DiRienzo-Zurbenko algorithm smoothing to Neagu-Zurbenko algorithm smoothing, in practice. The article concludes by delineating next steps to establish the precision with which Kolmogorov-Zurbenko periodograms that use these dynamic smoothing algorithms estimate signal strength.

## 1. Development and Logic of Kolmogorov-Zurbenko Periodograms with Dynamic Smoothing

This section provides a brief summary of the mathematical development and underlying logic of Kolmogorov-Zurbenko periodograms with dynamic smoothing. As stated in the Introduction, dynamic smoothing methods include the DiRienzo-Zurbenko algorithm (DiRienzo, 1996; DiRienzo & Zurbenko, 1999) and the Neagu Zurbenko algorithm (Neagu & Zurbenko, 2003; Zurbenko, 2001). Although standard periodograms are framed in terms of raw signal strength at each given frequency, time series analysts are often more comfortable framing spectral analyses in terms of information (i.e., entropy) by taking the log of a periodogram. As noted by Priestley (1981), using log-periodograms has distinct advantages over the use of standard periodograms. First, logarithmic transformation of a periodogram is a variance stabilizing technique because, unlike standard periodograms, the variance of its signal strength estimates is independent of frequency. Second, variance stabilization brings the distribution of signal strength estimates closer to normality. Third, comparisons across log-periodograms are independent of scaling units because differences are in percentages (or proportions) rather than the magnitudes of variance used in standard periodograms; as a result, it permits discussion of such differences in terms of universal units, such as decibels (dBs). For these reasons, Kolmogorov-Zurbenko periodograms are, by default, log-periodograms.

### 1.1 DiRienzo-Zurbenko Algorithm Smoothing

In approaching the problem of finding the spectral density and related smoothed periodogram, DiRienzo and Zurbenko (DiRienzo, 1996; DiRienzo & Zurbenko, 1999) began with a 'worst case scenario' by assuming a nonstationary time series process with a mixed spectrum. Defined in trigonometric form, it is:

$$X_t = \sum_{j=1}^{J} \rho_j \cos(\omega_j t + \phi_j) + \varepsilon_t, \qquad t = 0, \pm 1, \pm 2, \dots, \tag{1.1.1}$$

where $\omega_j$ is a constant frequency, $\rho_j$ is a constant amplitude, $\phi_j$ is a uniform phase shift on the interval $(-\pi, \pi)$, and $\varepsilon_t$ is random noise that is stationary and bounded with continuous spectral density $f_\varepsilon(\cdot)$. Consequently, the spectral density of $X_t$, that is $f_X(\cdot)$, is superimposed on the spectral density of $\varepsilon_t$, that is $f_\varepsilon(\cdot)$. However, the sample spectral density (i.e., periodogram) often has a considerable amount of variation, leading to a spectral estimate that is unstable and, thus, jagged, especially as sample size increases. To correct this, variation in the sample spectrum can be reduced by smoothing (Wei, 2006).

For an ordinary stationary process $X_t$, the smoothed spectral density is a function of a Fourier transform of the time series dataset, a selected spectral window form, and a selected spectral window bandwidth. One such basic estimate of the spectral density and its corresponding smoothed periodogram is the Grenander-Rosenblatt spectral estimate (Grenander, 1984), which is given by:

$$\hat{f}_N(\lambda_k) = \sum_{j=-N/2+1}^{N/2} \Phi_N(\lambda_j - \lambda_k) I_N(\lambda_j) \tag{1.1.2}$$

with periodogram

$$I_N(\lambda_k) = \frac{1}{2\pi N} |\sum_{t=0}^{N-1} X_t \exp(-i\lambda_k t)|^2 \quad (1.1.3)$$

calculated at the Fourier frequencies $\lambda_k = \frac{2\pi k}{N}$, where $k = -N/2 + 1, \dots, N/2$, and with $\Phi_N(\cdot)$ being a spectral window such that

$$\Phi_N(x) = A_N G(A_N x) \left\{ \int_{-A_N \pi}^{A_N \pi} G(\tilde{x}) d\tilde{x} \right\}^{-1}, \quad -\pi \le x \le \pi; \; 1 \ll A_N \ll N \quad (1.1.4)$$

where $G(\cdot)$ is the spectral window form and $A_N$ is the inverse of the spectral window bandwidth. Spectral window forms include uniform, Bartlett, Gaussian, Parzen, and Tukey-Hamming (Priestley, 1981; Zurbenko, 1986) and the spectral window width is assumed to be a selected constant.

However, the spectral estimate $\hat{f}_N(\lambda_k)$, can have a large bias, a large variance, or both; such a situation typically occurs when the underlying spectral density has a large amount of variability in its values or order of smoothness. Unfortunately, regular statistical estimations of spectra use fixed spectral window smoothing across the spectrum and this use of a fixed spectral window frequently smooths out some of the spectral lines; thus, the analyst is "... forced to compromise between variance reduction and bias" (Wei, 2006, p. 302)

To address this shortcoming, DiRienzo and Zurbenko developed a dynamic approach. First, they recognized that because the mean square error of a periodogram includes both bias and variance, it can be used to assess the quality of a spectral estimate, thus serving as the criterion for selecting the best spectral window and, further, that the mean square error can be used to assess the *local* quality of a spectral estimate (i.e., the quality around a given frequency). They also noted that optimal spectral estimation requires that both the window form and the bandwidth vary with frequency, but that bandwidth has a substantially greater impact on the asymptotic mean square error than does the window form. Consequently, DiRienzo and Zurbenko's dynamic approach ensures the optimal local quality of a spectrum estimate by focusing on selection of optimal *local* bandwidths.

To this end, they first proved two theorems. Their first theorem states that the asymptote of the local squared variation of $f(\lambda)$ is similar to the asymptote of the inverse of the optimal bandwidth, $A_N(\lambda)$, as defined in Equation (1.1.4). That is

$$\frac{T_\lambda^\Delta(f)}{C_{\lambda\Delta}} \approx N^{1/(1+2\alpha_\lambda)}(1-\epsilon) \quad (1.1.5)$$

where:

$T_\lambda^\Delta(f)$ = largest possible local squared variation;

$N^{1/(1+2\alpha_\lambda)}(1-\epsilon)$ = the optimal inverse bandwidth, $A_N(\lambda)$; and

$C_{\lambda\Delta}$ = constant under Hölder's condition.

Their second theorem states that the local squared variation of $f(\lambda)$ within the optimal bandwidth, $A_N^{-1}(\lambda)$, is virtually constant, independent of sample size. That is:

$$T_\lambda^{A_N^{-1}}(f) = \mathrm{K}_\lambda(1-\epsilon) \quad (1.1.6)$$

where:

$T_\lambda^{A_N^{-1}}(f)$ = local squared variance for spectral window bandwidth $A_N^{-1}(x)$;

$A_N^{-1}(\lambda)$ = the optimal spectral window bandwidth, with $\Delta > A_N^{-1} > 0$; and

$\mathrm{K}_\lambda = \frac{C_\lambda^2 (2\pi)^{2\alpha_\lambda}}{C_\lambda^{opt} \pi}$, where $2\alpha_\lambda$ is the window width, $C_\lambda$ is a spectral constant under Hölder's condition, and $C_\lambda^{opt}$ is the optimal spectral constant under Hölder's condition.

Based on these two theorems, they created an algorithm which permits a dynamic (i.e., variable) bandwidth that depends on the local characteristics of the underlying spectral density.

The logic behind the DiRienzo-Zurbenko algorithm is as follows. For each frequency, $\lambda_k$, in Equation (1.1.2), the width of the spectral window is increased until the local squared variation of the periodogram, $I_N(\cdot)$, reaches a pre-selected constant value; this constant value is selected by the analyst and is a set proportion of the total variance in the periodogram (e.g., 0.05, 0.01), designated as the *DiRienzo-Zurbenko proportion of smoothness*. Consequently, the resulting smoothed periodogram is formed using a dynamic window – when a signal is present, local variance is high, the width of the window becomes narrow (i.e., zooms in), and the smoothed periodogram spikes up; conversely when no signal is present, local variance is low, the width of the window becomes wide (i.e., zooms out), and the smoothed periodogram becomes relatively flat. As a result, the Kolmogorov-Zurbenko periodogram with DiRienzo-Zurbenko algorithm smoothing more clearly displays a spectral estimate that is sensitive to weak signals, accurate in identifying frequencies, and can resolves signals that are close in frequency.

## 1.2 Neagu-Zurbenko Algorithm Smoothing

An alternative algorithm that dynamically smooths log-periodograms has also been developed. Based on Kolmogorov's proof that the error of prediction for a stationary sequence is constant (Kolmogorov, 1941, 1992), Kolmogorov's algorithmic approach to quantifying information (Kolmogorov, 1965), and Wu's casting the spectral domain as information (Wu, 1997), Zurbenko (2001) developed a second algorithm that also dynamically smooths log-periodograms, but in a different way: in contrast to the DiRienzo-Zurbenko algorithm which dynamically smooths a log-periodogram by focusing on local changes in variance, the Neagu-Zurbenko algorithm dynamically smooths a log-periodogram by focusing on local changes in its linear rate of change. Consequently, the Neagu-Zurbenko algorithm has two distinct advantages over the DiRienzo-Zurbenko algorithm. First, it eliminates spurious spectral peaks in the periodogram and, second, it is better able to resolve close but unequal spectral peaks (Neagu & Zurbenko, 2003; Zurbenko, 2001).

Defining the log spectral density estimate as a local average of the log-periodogram

$$L_a(\lambda) = \frac{1}{2\cdot a(\lambda)} \int_{-a(\lambda)}^{a(\lambda)} \log I_n(\lambda + \omega) d\omega \text{, with } -\frac{1}{2} < \lambda \le \frac{1}{2} \quad (1.2.1)$$

where $I_n(\cdot)$ is the periodogram and $2 \cdot a(\cdot)$ is the size of the window, Zurbenko proved that the *estimated* spectral density that minimizes the cross entropy with the *true* spectral density is defined by the spectral window width $2 \cdot a(\lambda)$, and that the window width has the asymptotic property that the information in the process changes at a linear rate for that given location in the spectra (i.e., there is a boundary beyond which change is no longer linear). That is, for a given frequency, $\lambda$,

$$\int_{-a(\lambda)}^{a(\lambda)} \log f(\lambda + \omega) d\omega = 2 \cdot a(\lambda) \log f(x) \quad (1.2.2)$$

where:

$f(\cdot)$ = spectral density; and

$2 \cdot a(\lambda)$ = window width.

Further, this holds over the whole domain of $\lambda$ because, as Zurbenko subsequently proved, the correspondence of $\lambda \rightarrow a_\lambda$ as a function from $(-\frac{1}{2}, \frac{1}{2}] \rightarrow [0, \frac{1}{2}]$ is continuous, meaning that one can integrate over $\lambda$ and obtain the minimum of the integral. Because the total departure from linearity in a given spectrum is a fixed value, the analyst can set a proportion of the total departure from linearity found in the spectrum. In this way, the window width is dynamic, with $2 \cdot a(\lambda)$ relatively wide when the local spectrum is flat and relatively narrow when the local spectrum spikes up, indicative of departure from linearity and, in turn, indicative that a signal is present at that frequency.

Zurbenko's alternative approach was subsequently refined by Neagu and Zurbenko to form the Neagu-Zurbenko algorithm (Neagu & Zurbenko, 2003). Like the DiRienzo-Zurbenko algorithm, it smooths the log-periodogram using a dynamic window, but this alternate approach first computes a linear approximation to the local information in the process, then uses the dynamic window to minimize the dissimilarity between the estimate and the log of the observed spectrum.

As a result, the Kolmogorov-Zurbenko periodogram with Neagu-Zurbenko smoothing utilizes a dynamic window that expands and contracts based on a set proportion of the total departure from linearity found in the given spectrum (e.g., 0.05, 0.01), designated as the *Neagu-Zurbenko proportion of smoothness*. Thus, it 'zooms in' when local information spikes up relative to the linear approximation, indicating the presence of a signal, and it 'zooms out' when local information is low and commensurate with the linear estimate, indicating that no signal is present. While a preliminary comparison of the Kolmogorov-Zurbenko periodogram with DiRienzo-Zurbenko algorithm smoothing to one using Neagu-Zurbenko algorithm smoothing can be found in Neagu and Zurbenko (2003), the current investigation provides a more detailed comparison of these two smoothing algorithms with respect to the theoretical and practical limits of their sensitivity, accuracy, resolution, and robustness in estimating signal frequencies.

## 2. Access and Utilization of Software

While DiRienzo and Zurbenko as well as Neagu and Zurbenko have fully described the motivation, development, and basic principles underlying their respective algorithms, there is little in the way of how one can access and utilize them. This section provides information about where the DiRienzo-Zurbenko and Neagu-Zurbenko algorithms can be found and how to use them with a given dataset.

### 2.1 Access

The DiRienzo-Zurbenko and Neagu-Zurbenko algorithms are available in the *R* platform for data analysis and are options in the *kzp( )* function within the *kza* package at https://CRAN.R-project.org/package=kza (Close et al., 2020). Because errors can occur when downloading the *kza* package using *R*'s *install.packages( )* command, the analyst should download the *kza* Package Source folder using the *kza_4.1.01.tar.gz* link; within the *kza* folder is an *R* sub-folder that includes three *R* script files: *rlv.R*, *kza.R*, and *kzft.R*. The *kzp( )* function is in the *kzft.R* script file. However, before conducting a time series analysis, the analyst should delete an incorrect version of the *plot.kzp( )* function in the *kza.R* script, thus retaining the correct version of the *plot.kzp( )* function placed in the *kzft.R* script. Next, the analyst should run the *kza.R* script, the *kzft.R* script, and the *rlv.R* script, in that order, to ensure any dependencies among and between scripts are addressed. To summarize:

Go to: https://CRAN.R-project.org/package=kza
Download the *kza* folder by clicking the *kza_4.1.0.1.tar.gz* link
Open the *kza* folder
Open the *R* sub-folder
Delete the incorrect version of the *plot.kzp( )* function in the *kza.R* script
Run the edited *kza.R* script file, the *kzft.R* script file, and the *rlv.R* script file, in that order.

As discussed below, the DiRienzo-Zurbenko and Neagu-Zurbenko algorithms are options in the *kzp( )* function found in the *kzft.R* script file. Thus, the *kzp( )* function is used to construct the Kolmogorov-Zurbenko periodogram for a given time series dataset and results in a spectral density estimate using the Kolmogorov-Zurbenko Fourier transform (KZFT). In turn, the DiRienzo-Zurbenko and Neagu-Zurbenko algorithms are optional methods that are accessed when running the *kzp( )* function.

### 3.2 Utilization

To construct a Kolmogorov-Zurbenko periodogram utilizing either the DiRienzo-Zurbenko algorithm or the Neagu-Zurbenko algorithm for a given time series dataset, a researcher must specify the raw time series dataset to be analyzed, the initial width of the filtering window, the number of iterations of the KZFT, the smoothness level of the periodogram, the DiRienzo-Zurbenko algorithm or the Neagu-Zurbenko algorithm as the method for smoothing, and the number of top signal frequencies and periods to be identified. With respect to the number of iterations, the first iteration applies the KZFT to the time series dataset, the second iteration applies the KZFT to the result of the first iteration, and so on, such that the transfer function for the selected number of iterations is a product of the selected number of iterations of the KZFT transfer function. With a starting frequency of 0, one iteration results in a standard periodogram while two or more iterations approach a Gaussian distribution, thereby eliminating false low-level frequency spikes.

In its basic form, generating a Kolmogorov-Zurbenko periodogram with either DiRienzo-Zurbenko algorithm or Neagu-Zurbenko algorithm smoothing utilizes four functions:

**totalSignal <- kzp(y, m, k)**
**kzPeriodogram <- smooth.kzp(totalSignal, log=TRUE, smooth.level, method)**
**kzpPlot <- plot.kzp(kzPeriodogram, type = "1")**
**kzpSummary <- summary.kzp(kzPeriodogram, digits, top)**

where the arguments are:

| | |
|---|---|
| y | raw time series dataset (a time ordered vector of values) |
| m | initial window width, $m_{Initial}$ |
| k | number of iterations of KZFT |
| smooth_level | proportion of total variance (for method = "DZ") *or*<br>proportion of total departure from linearity (for method = "NZ")<br>(a.k.a. proportion of smoothness) |
| method | method for smoothing:<br>"DZ" for DiRienzo-Zurbenko, "NZ" for Neagu-Zurbenko |
| digits | number of significant digits |
| top | number of top frequencies and periods to be reported |

As a result, the *kzp( )* function uses the Kolmogorov-Zurbenko Fourier transform to generate the raw periodogram; *smooth.kzp( )* returns the smoothed log-periodogram ordinates and the smoothing method; *plot.kzp( )* returns a plot of the smoothed Kolmogorov-Zurbenko periodogram that is *mean-centered* (i.e., the mean of all ordinate values); and *summary.kzp( )* returns the *top* frequencies and *top* periods of interest.

In this basic form, the *kzp( )* function estimates the spectral density across the range of frequencies through use of a moving average. The initial width of the window used to compute the moving average is selected by the researcher (i.e., m=$m_{Initial}$). When the *smooth.kzp( )* function is invoked, the DiRienzo-Zurbenko algorithm or the Neagu-Zurbenko algorithm adapts the width in such a way as to 'zoom in' where values of the spectral density spike up and 'zoom out' where values of the spectral density are relatively flat. It does this by utilizing the preselected smoothness level; that is, the selected smoothness level is a proportion of total variance in the periodogram for the DiRienzo-Zurbenko algorithm or proportion of total departure from linearity in the periodogram for the Neagu-Zurbenko algorithm. With the smoothness level set, the width of the window used in computing moving averages across the frequency spectrum varies so that the proportion of variance or proportion departure from linearity contained in the window remains constant (i.e., the set proportion); thus, the window shrinks in size when the spectral density spikes up (i.e., local variance or local departure from linearity is relatively high) and expands in size when spectral density is relatively flat (i.e., local variance or local departure from linearity is relatively low). In turn, the *plot.kzp( )* function generates a plot of the smoothed Kolmogorov-Zurbenko periodogram and the *summary.kzp( )* function returns the *top* frequencies and *top* periods in the time series dataset.

## 3. Theoretical and Practical Limits

As noted above, the purpose of this study is to establish the theoretical and practical limits of Kolmogorov-Zurbenko periodograms with dynamic smoothing with respect to sensitivity, accuracy, resolution, and robustness in estimating signal frequencies, and to compare its predictive power to that of a standard ARIMA model. Theoretical and practical limits are established in this Section, while predictive power is demonstrated in Section 4.

Establishing the limits of Kolmogorov-Zurbenko periodograms with dynamic smoothing in estimating signal frequencies utilize four criteria: (1) sensitivity, (2) accuracy, (3) resolution, and (4) robustness with respect to missing data. For each criterion, this section provides a brief definition, importance, statistical bases, theoretical limits where available, practical limits where applicable, and demonstrations of Kolmogorov-Zurbenko periodograms with dynamic smoothing performance across a series of simulated datasets.

All demonstrations used signals at Fourier frequencies consistent with selected algorithm settings. Initial window widths were set to $m_{Initial} = 500$ and the number of iterations of the KZFT was set to $k = 3$. For each criterion, two different numbers of observations were examined, $n = 5000$ and $n = 1000$, and two different levels for the proportion of smoothness were utilized, either $DZ = 0.05$ and $DZ = 0.01$ or $NZ = 0.05$ and $NZ = 0.01$; taken in combination, this resulted in four possible scenarios across each of the four criteria (i.e., sixteen series of analyses) for each dynamic smoothing algorithm.

### 3.1 Sensitivity

*Definition and Importance*. The first criterion in assessing the performance of a given periodogram algorithm is sensitivity. In spectral analysis, *sensitivity* is the ability to detect a signal embedded in random noise. Thus, a smoothing algorithm can be assessed with respect to the signal-to-noise ratio present in a given time series dataset. As such, the analyst must know how strong a given signal must be relative to the strength of the surrounding noise in order for it to be detected. To this end, a good smoothing algorithm will be able to detect a weak signal in the presence of high levels of noise. Without this quality, an algorithm will be virtually useless with all but the strongest of signals relative to noise.

*Statistical Basis and Limits*. With regard to the theoretical limit for sensitivity using the Kolmogorov-Zurbenko periodogram with DiRienzo-Zurbenko algorithm smoothing, the signal-to-noise ratio must be greater than or equal to the proportion of DiRienzo-Zurbenko smoothing set by the analyst. The total variance of a Kolmogorov-Zurbenko periodogram consists of variance due to one or more signals, $s^2$, and variance due to noise, $n^2$. In order for a signal to be detected using the DiRienzo-Zurbenko algorithm, the variance of a signal must be greater than or equal to the variance due to noise in the locale of that signal frequency. When using the DiRienzo-Zurbenko algorithm, the amount of total variance in the locale of a signal frequency is restricted to the proportion of total variance, also known as the proportion of smoothing, $DZ$, as set by the analyst. Thus, the variance of a signal at a given frequency must be greater than or equal to the proportion of variance of noise in the locale of that frequency. That is:

$$s^2 \geq DZ \cdot n^2 \qquad (3.1.1)$$

where:

$s^2$ = variance of signal;
$n^2$ = total variance of noise; and
$DZ$ = DiRienzo-Zurbenko proportion of smoothness

Indeed, stated differently, this becomes:

$$\frac{s^2}{n^2} \geq DZ \qquad (3.1.2)$$

Because $\frac{s^2}{n^2}$ is none other than the signal-to-noise ratio, the limit for detecting a signal is a signal-to-noise ratio equal to the DiRienzo-Zurbenko proportion of smoothness. Typical values for the DiRienzo-Zurbenko proportion of smoothness are 0.05 and 0.01. Thus, the theoretical limit of the signal-to-noise ratio for detecting a signal is quite small.

As just noted, the DiRienzo-Zurbenko algorithm sets the proportion of smoothness in terms of total variance in the periodogram and, as a result, its theoretical limit is in terms of the signal-to-noise ratio. However, the Neagu-Zurbenko algorithm sets the proportion of smoothness in terms of the total departure from linearity in a periodogram and, as a result, it does not have a theoretical limit in terms of the signal-to-noise ratio. Nevertheless, we can demonstrate the performance of the Neagu-Zurbenko algorithm with respect to the signal-to-noise ratio by evaluating its performance over a range of signal-to-noise ratios that spans the theoretical limit of the DiRienzo Zurbenko algorithm. Accordingly, this allows for the possibility that the sensitivity of the Neagu-Zurbenko surpasses the sensitivity of the DiRienzo-Zurbenko algorithm.

With standard periodograms using static smoothing with fixed window width, the Rose criterion (Rose, 1973) indicates a lower bound of $s/n = 3$, or $s^2/n^2 = 9$,. Consequently, the Kolmogorov-Zurbenko periodogram with DiRienzo-Zurbenko algorithm smoothing has the potential to detect considerably weaker signals embedded in high levels of noise.

*Demonstration*. To demonstrate sensitivity, the amplitude of noise was set at 16 and the signal-to-noise ratio was made to vary by adjusting the amplitude of the respective signal across the four scenarios. For scenarios in which the DiRienzo-Zurbenko proportion of smoothness was set at 0.05, the signal-to-noise ratio ranged from 0.057 to 0.050, (i.e., its theoretical limit), in increments of 0.001. Likewise, for scenarios in which the DiRienzo-Zurbenko proportion of smoothness was set at 0.01, the signal-to-noise ratio ranged from 0.017 to 0.010 (i.e., its theoretical limit), in increments of 0.001. For scenarios in which the Neagu-Zurbenko proportion of smoothness was set at 0.05, the signal-to-noise ratio ranged from 0.057 to 0.048 in increments of 0.001, notably spanning the theoretical limit of 0.050 for the DiRienzo-Zurbenko algorithm. Likewise, for scenarios in which the proportion of smoothness was set at 0.01, the signal-to-noise ratio ranged from 0.017 to 0.008, in increments of 0.001, again notably spanning the theoretical limit of 0.010 for the DiRienzo-Zurbenko algorithm. The ranges of signal-to-noise ratio for the two levels of smoothness (i.e., 0.05, 0.01) are different because the theoretical limit of sensitivity of the DiRienzo-Zurbenko algorithm for a given scenario is determined by the level of the proportion of smoothness in terms of variance.

In scenarios where the number of observations was set at $n = 5000$, the Kolmogorov-Zurbenko periodogram with DiRienzo-Zurbenko algorithm smoothing reached the respective theoretical limits in terms of signal-to-noise ratio, detecting the signal frequency $\lambda$ = 0.040, regardless of proportion of smoothness. When the number of observations was set to $n = 1000$ and proportion of smoothness set at *DZ* = 0.05, a signal frequency of 0.044, rather than 0.040, was detected across its entire range of signal-to-noise ratios. However, with $n = 1000$ and proportion of smoothness set at *DZ* = 0.01, the theoretical limit with respect to signal-to-noise ratio was approached: the signal frequency of 0.040 was detected up to a signal-to-noise ratio of 0.013, but the signal was lost when moving from a signal-to-noise ratio of 0.013 to one of 0.012. Thus, it appears the Kolmogorov-Zurbenko periodogram with DiRienzo-Zurbenko algorithm smoothing using relatively large proportions of smoothness (e.g., 0.05) becomes less sensitive as number of observations decreases and the sensitivity of Kolmogorov-Zurbenko periodogram with DiRienzo-Zurbenko algorithm smoothing using relatively small proportions of smoothness (e.g., 0.01) approaches but does not achieve its theoretical limit as the number of observations decreases (see Table 3.1.1.).

In scenarios where the number of observations was set at $n = 5000$, the Kolmogorov-Zurbenko periodogram with Neagu-Zurbenko algorithm smoothing surpassed the respective theoretical limits of the DiRienzo-Zurbenko algorithm in terms of signal-to-noise ratio, detecting the signal frequency $\lambda = 0.040$, regardless of proportion of smoothness (i.e., *NZ* = 0.05, *NZ* = 0.01). When the number of observations was set to $n = 1000$ and proportion of smoothness set at *NZ* = 0.05, a signal frequency of 0.044, rather than 0.040, was detected across its entire range of signal-to-noise ratios. However, with $n = 1000$ and proportion of smoothness set at *NZ* = 0.01, the theoretical limit of the DiRienzo-Zurbenko algorithm with respect to signal-to-noise ratio was approached: the signal frequency of 0.040 was detected up to a signal-to-noise ratio of 0.011, but the signal was lost when moving to the DiRienzo-Zurbenko algorithm theoretical limit of a signal-to-noise ratio of 0.010. (See Table 3.1.2.).

Thus, it appears the Kolmogorov-Zurbenko periodogram with Neagu-Zurbenko algorithm smoothing using relatively large proportions of smoothness (e.g., 0.05) becomes less sensitive as the number of observations decreases and the sensitivity of Kolmogorov-Zurbenko periodogram with Neagu-Zurbenko algorithm smoothing using relatively small proportions of smoothness (e.g., 0.01) all but achieves the DiRienzo-Zurbenko algorithm theoretical limit as the number of observations decreases. Given that (1) the Neagu-Zurbenko algorithm surpasses the DiRienzo-Zurbenko algorithm practical limit for sensitivity when $n$ is relatively large ( $n = 5000$) and (2) the Neagu-Zurbenko algorithm more closely approaches the practical limit of the DiRienzo-Zurbenko algorithm than the DiRienzo-Zurbenko algorithm did itself when $n$ is relatively small ($n = 1000$) and a smaller proportion of smoothness is invoked, the Neagu-Zurbenko algorithm appears to perform somewhat better than the DiRienzo-Zurbenko algorithm with respect to sensitivity.

Based on these findings, subsequent demonstrations for accuracy, resolution, and robustness set the respective signal-to-noise ratio at 0.10. While this value is above the theoretical limit for both $DZ = 0.05$ and $DZ = 0.01$, it is well below the accepted lower bound of 9 using the Rose criterion.

| **SENSITIVITY: OBSERVED TOP SIGNAL FREQUENCY**<br>**Actual Signal Frequency = 0.040**<br>**Amplitude of Noise = 16** | | | | | | | |
|---|---|---|---|---|---|---|---|
| $n = 5000$ | | | | $n = 1000$ | | | |
| ***DZ* = 0.05** | | ***DZ* = 0.01** | | ***DZ* = 0.05** | | ***DZ* = 0.01** | |
| $\frac{s^2}{n^2}$ | **Observed Top Frequency** | $\frac{s^2}{n^2}$ | **Observed Top Frequency** | $\frac{s^2}{n^2}$ | **Observed Top Frequency** | $\frac{s^2}{n^2}$ | **Observed Top Frequency** |
| 0.057 | 0.040 | 0.017 | 0.040 | 0.057 | 0.044 | 0.017 | 0.040 |
| 0.056 | 0.040 | 0.016 | 0.040 | 0.056 | 0.044 | 0.016 | 0.040 |
| 0.055 | 0.040 | 0.015 | 0.040 | 0.055 | 0.044 | 0.015 | 0.040 |
| 0.054 | 0.040 | 0.014 | 0.040 | 0.054 | 0.044 | 0.014 | 0.040 |
| 0.053 | 0.040 | 0.013 | 0.040 | 0.053 | 0.044 | 0.013 | 0.040 |
| 0.052 | 0.040 | 0.012 | 0.040 | 0.052 | 0.044 | 0.012 | 0.160 |
| 0.051 | 0.040 | 0.011 | 0.040 | 0.051 | 0.044 | 0.011 | 0.160 |
| 0.050 | 0.040 | 0.010 | 0.040 | 0.050 | 0.044 | 0.010 | 0.160 |

**Table 3.1.1. Observed Top Frequencies in Kolmogorov-Zurbenko Periodograms using DiRienzo-Zurbenko Algorithm Smoothing across Varying Signal-To-Noise Ratios for Signal Frequency = 0.040 and Noise Amplitude = 16, with Number of Observations $n = 5000$ or $n = 1000$.**
**NB. Initial Window Width $m_{Initial} = 500$, Number of Iterations $k = 3$, and Proportions of Smoothness $DZ = 0.05$ and $DZ = 0.01$.**

| **SENSITIVITY: OBSERVED TOP SIGNAL FREQUENCY**<br>**Actual Signal Frequency = 0.040**<br>**Amplitude of Noise = 16** | | | | | | | |
|---|---|---|---|---|---|---|---|
| $n = 5000$ | | | | $n = 1000$ | | | |
| $NZ = 0.05$ | | $NZ = 0.01$ | | $NZ = 0.05$ | | $NZ = 0.01$ | |
| $\frac{s^2}{n^2}$ | **Observed Top Frequency** | $\frac{s^2}{n^2}$ | **Observed Top Frequency** | $\frac{s^2}{n^2}$ | **Observed Top Frequency** | $\frac{s^2}{n^2}$ | **Observed Top Frequency** |
| 0.057 | 0.040 | 0.017 | 0.040 | 0.057 | 0.044 | 0.017 | 0.040 |
| 0.056 | 0.040 | 0.016 | 0.040 | 0.056 | 0.044 | 0.016 | 0.040 |
| 0.055 | 0.040 | 0.015 | 0.040 | 0.055 | 0.044 | 0.015 | 0.040 |
| 0.054 | 0.040 | 0.014 | 0.040 | 0.054 | 0.044 | 0.014 | 0.040 |
| 0.053 | 0.040 | 0.013 | 0.040 | 0.053 | 0.044 | 0.013 | 0.040 |
| 0.052 | 0.040 | 0.012 | 0.040 | 0.052 | 0.044 | 0.012 | 0.040 |
| 0.051 | 0.040 | 0.011 | 0.040 | 0.051 | 0.044 | 0.011 | 0.040 |
| 0.050 | 0.040 | 0.010 | 0.040 | 0.050 | 0.044 | 0.010 | 0.160 |
| 0.049 | 0.040 | 0.009 | 0.040 | 0.049 | 0.044 | 0.009 | 0.160 |
| 0.048 | 0.040 | 0.008 | 0.040 | 0.048 | 0.044 | 0.008 | 0.160 |

**Table 3.1.2. Observed Top Frequencies in Kolmogorov-Zurbenko Periodograms using Neagu-Zurbenko Algorithm Smoothing across Varying Signal-To-Noise Ratios for Signal Frequency = 0.040 and Noise Amplitude = 16, with Number of Observations $n = 5000$ or $n = 1000$.**
**NB. Initial Window Width $m_{Initial} = 500$, Number of Iterations $k = 3$, and Proportions of Smoothness $NZ = 0.05$ and $NZ = 0.01$.**

**3.2 Accuracy**

*Definition and Importance*. The second criterion in assessing the performance of a given periodogram algorithm is accuracy. In spectral analysis, *accuracy* is the ability to precisely identify the frequency of a given signal. Combined with sensitivity, accuracy ensures a periodogram can pinpoint the exact frequency of a weak signal embedded in high levels of noise. Without accuracy, predicted values of the variable will quickly diverge from the true values as one moves forward in time; conversely, with high levels of accuracy, predicted values of the variable will remain close to the true values into perpetuity for all intents and purposes.

*Statistical Basis and Limits*. For Kolmogorov-Zurbenko periodograms with dynamic smoothing across the frequency spectrum from $-\pi$ to $+\pi$, the theoretical limit for accuracy is $\pi/n$. Kolmogorov-Zurbenko periodograms are determined by using the Kolmogorov-Zurbenko Fourier transform (KZFT). The KZFT works by reconstructing the original signal in a raw time series dataset using a weighted sum of all possible empirical sinusoidal waves that could be present in the dataset. Because there are $n$ observations in a given dataset, there are $n-1$ possible sinusoidal waves, with any two consecutive waves separated by $\pi/n$. Given that deviation of a signal by more than $\pi/(2n)$ above or below the signal frequency would be attributed to the next possible sinusoidal wave for the given number of observations, the theoretical limit of accuracy for the frequency of any given signal is $\pi/n$.

At the theoretical limits, the accuracy of dynamic smoothing windows is equivalent to or surpasses that of static smoothing windows. For a given static smoothing window, its accuracy is equal to its bandwidth. Typical static smoothing windows and their respective estimated bandwidths are: rectangular with $\pi/M$, Tukey-Hamming with $2.5\pi/M$, Tukey-Hanning with $2.57\pi/M$, Bartlet with $3\pi/M$, and Parzen with $3.7\pi/M$ (Koopmans, 1995). Here, $M$ is the selected truncation point for the static smoothing window and it has an upper limit of $n-1$ (Wei, 2006). As a result, the theoretical limit of accuracy for each of these static smoothing windows is: rectangular with $\pi/(n-1)$, Tukey-Hamming with $2.5\pi/(n-1)$, Tukey-Hanning with $2.57\pi/(n-1)$, Bartlett with $3\pi/(n-1)$, and Parzen with $3.7\pi/(n-1)$. Each of these is equal to or larger than the theoretical limit for accuracy of dynamic smoothing windows.

In practice, dynamic smoothing algorithms identify signal frequencies by examining the local quality of the spectral estimate using a dynamic window with an initial window with of $m_{Initial}$. Thus, the practical limit of accuracy for Kolmogorov-Zurbenko periodograms with dynamic smoothing is $\pi/m_{Initial}$. It should be noted that the maximum initial window width is $n$, so both the theoretical limit and the smallest possible practical limit of accuracy is the same, $\pi/n$.

*Demonstration*. To demonstrate accuracy, the amplitude of noise was set at 16 and the signal-to-noise ratio was set at 0.10, resulting in signal amplitudes of 5.06. Actual signal frequencies were set at 0.400, 0.440, or 0.444, in an effort to demonstrate accuracy in tenths, hundredths, and thousands. As with demonstration of sensitivity, two numbers of observations were considered, *n* = 5000 and *n* = 1000, as were two levels of proportion of smoothness, either *DZ* = 0.05 and *DZ* = 0.01 or *NZ* = 0.05 and *NZ* = 0.01. The initial window width was set at $m_{Initial} = 500$, resulting in a practical limit for accuracy of $0.002$.

For DiRienzo-Zurbenko algorithm smoothing, signal frequency estimates were accurate to the thousands across all four scenarios: $n = 5000, DZ = 0.05$; $n = 5000, DZ = 0.01$; $n = 1000, DZ = 0.05$; and $n = 1000, DZ = 0.01$. Thus, accuracy was maintained regardless of whether the number of observations was relatively large or small and regardless of whether the proportion of smoothness was relatively high or low. (See Table 3.2.1.)

In scenarios where the number of observations was set at $n = 1000$, the Kolmogorov-Zurbenko periodogram with Neagu-Zurbenko algorithm smoothing was able to accurately identify signal frequencies within the practical limit of $0.002$ regardless of whether the proportion of smoothness was set to *NZ* = 0.05 or to *NZ* = 0.01. With $n = 1000$ and *NZ* = 0.05, none of the frequencies were accurately identified. Nevertheless, when the number of observations was *n* = 1000 with *NZ* = 0.01, it was able to identify the signal frequencies of 0.400, 0.440, and 0.444 exactly, thus within the practical limit of 0.002. Consequently, it appears that accuracy using relatively high proportions of smoothness (e.g., 0.05) decreases as the number of observations decreases, while accuracy using relatively low proportions of smoothness (e.g., 0.01) is within the practical limit regardless of whether the number of observations is relatively large (e.g., $n = 5000$) or relatively small (e.g., $n = 1000$). Finally, with the notable exception of the scenario with a small number of observations ($n = 1000$) and a high proportion of smoothness ($NZ = 0.05$) where the Neagu-Zurbenko algorithm performed worse, these results duplicate the results obtained in the demonstration of accuracy for the Kolmogorov-Zurbenko periodogram with DiRienzo-Zurbenko algorithm smoothing. (See Table 3.2.2.)

| **ACCURACY: OBSERVED TOP SIGNAL FREQUENCY**<br>**Amplitude of Signal = 5.06**<br>**Amplitude of Noise = 16**<br>**Signal-to-Noise Ratio = 0.10** | | | | |
|---|---|---|---|---|
| | $\boldsymbol{n = 5000}$ | | $\boldsymbol{n = 1000}$ | |
| **Actual Signal Frequency** | ***DZ* = 0.05** | ***DZ* = 0.01** | ***DZ* = 0.05** | ***DZ* = 0.01** |
| 0.400 | 0.400 | 0.400 | 0.400 | 0.400 |
| 0.440 | 0.440 | 0.440 | 0.440 | 0.440 |
| 0.444 | 0.444 | 0.444 | 0.444 | 0.444 |

**Table 3.2.1. Observed Top Frequencies in Kolmogorov-Zurbenko Periodograms using DiRienzo-Zurbenko Algorithm Smoothing across Varying Actual Signal Frequencies, with Signal-to-Noise Ratio = 0.10 and Number of Observations $n = 1000$ or $n = 1000$.**
**NB. Initial Window Width $m_{Initial} = 500$, Number of Iterations $k = 3$, and Proportion of Smoothness *DZ* = 0.05 and *DZ* = 0.01.**

| **ACCURACY: OBSERVED TOP SIGNAL FREQUENCY<br>Amplitude of Signal = 5.06<br>Amplitude of Noise = 16<br>Signal-to-Noise Ratio = 0.10** | | | | |
|---|---|---|---|---|
| | $n = 5000$ | | $n = 1000$ | |
| **Actual Signal Frequency** | ***NZ* = 0.05** | ***NZ* = 0.01** | ***NZ* = 0.05** | ***NZ* = 0.01** |
| 0.400 | 0.400 | 0.400 | 0.404 | 0.400 |
| 0.440 | 0.440 | 0.440 | 0.444 | 0.440 |
| 0.444 | 0.444 | 0.444 | 0.448 | 0.444 |

**Table 3.2.2. Observed Top Frequencies in Kolmogorov-Zurbenko Periodograms using Neagu-Zurbenko Algorithm Smoothing across Varying Actual Signal Frequencies, with Signal-to-Noise Ratio = 0.10 and Number of Observations $n = 5000$ or $n = 1000$. NB. Initial Window Width $m$ = 500, Number of Iterations $k = 3$, and Proportion of Smoothness *NZ* = 0.05 and *NZ* = 0.01.**

**3.3 Resolution**
*Definition and Importance*. The third criterion in assessing the performance of a given periodogram algorithm is resolution. In spectral analysis, *resolution* is the ability to sensitively detect and accurately identify the frequencies of two signals that are close in frequency. In this situation, it is important to know how close two frequencies can be before a smoothing algorithm is no longer able to separate them. Without the ability to resolve close signals, a periodogram will merge the signals into a single peak, making accurate prediction of future values of a variable all but impossible. Thus, despite having a common starting point, two signals with two different frequencies will have a much different effect on the predicted value of a variable over time compared to one signal of a single frequency. Consequently, a smoothing algorithm must be able to not only identify each signal with accuracy but must also have the ability to resolve two such signals when they are close together.

*Statistical Basis and Limits*. For Kolmogorov-Zurbenko periodograms with dynamic smoothing across the frequency spectrum from $-\pi$ to $+\pi$, the theoretical limit for resolution is $2\pi/n$. As stated above, the Kolmogorov-Zurbenko periodogram is determined by using the KZFT which reconstructs an original raw dataset using a weighted sum of all possible empirical sinusoidal waves that could be present in a time series dataset. Given $n$ observations in such a dataset, there are $n-1$ possible sinusoidal waves and any two consecutive sinusoidal waves have a separation of $\pi/n$. Clearly, the spike in spectral density of two consecutive empirical frequencies would overlap and could not be distinguished from one another. However, if the distance between two close frequencies is $2 \cdot (\pi/n) = 2\pi/n$ or greater, then resolution of the frequencies becomes possible. Thus, the theoretical limit for resolution of two signals that are close to one another is a separation in frequency of $2\pi/n$.

At their theoretical limits, resolution of dynamic smoothing windows is equivalent to or surpasses that of static smoothing windows. For a given static smoothing window, its resolution is equivalent to twice its bandwidth. Given the bandwidth of typical static smoothing windows in the previous subsection on accuracy along with the truncation point's upper limit of $n-1$, the theoretical limit for resolution of each static smoothing window under consideration is: rectangular with $2 \cdot \pi/(n-1)$, Tukey-Hamming with $2 \cdot 2.5\pi/(n-1)$, Tukey-Hanning with $2 \cdot 2.57\pi/(n-1)$, Bartlett with $2 \cdot 3\pi/(n-1)$, and Parzen with $2 \cdot 3.7\pi/(n-1)$. Because each of these is equal to or larger than the theoretical limit for resolution of dynamic smoothing windows, $2\pi/n$, the resolution of dynamic smoothing windows is as good as or surpasses that of the static smoothing windows under consideration.

In practice and as with accuracy, dynamic smoothing algorithms identify signal frequencies by examining the local quality of the spectral estimate using a dynamic window with an initial window with of $m_{Initial}$. Thus, the practical limit of resolution for Kolmogorov-Zurbenko periodograms with dynamic smoothing is $2/m_{Initial}$. It should be noted that the maximum initial window width is $n$, so both the theoretical limit and the smallest possible practical limit of resolution is the same, $2/n$.

*Demonstration*. To demonstrate resolution, the amplitude of noise was set at 16 and the signal-to-noise ratio was set at 0.10, resulting in signal amplitudes of 3.58 for each of the two signals. The initial window width was set a $m_{Initial} = 500$, with a practical limit for accuracy of 0.002 and a practical limit for resolution of $0.004$. Because this was a test of how well the Kolmogorov-Zurbenko periodogram with dynamic smoothing can resolve two frequencies that are close to one another, one frequency was fixed at $\lambda_1 = 0.040$, while a second signal's frequency was increased so as to approach the first signal frequency up to the practical limit for resolution of 0.004, with $\lambda_2 = 0.030, 0.032, 0.034$, and $0.036$. Again, and as with the demonstration of sensitivity and accuracy, the numbers of observations were $n = 5000$ and $n = 1000$ and the proportions of smoothness were either *DZ* = 0.05 and *DZ* = 0.01 or *NZ* = 0.05 and *NZ* = 0.01.

With regard to resolution, Kolmogorov-Zurbenko periodograms with DiRienzo-Zurbenko algorithm smoothing successfully resolved signal pairs across scenarios with $n = 5000$, regardless of whether the proportion of smoothness was *DZ* = 0.05 or *DZ* = 0.01. In these scenarios, top observed frequencies were within the practical limit for accuracy (0.002) and, thus, the separate frequencies were accurate and clearly resolved. When the number of observations was reduced to $n = 1000$ with $DZ = 0.05$, performance of the DiRienzo-Zurbenko algorithm was poor: top signal frequencies were within 0.002 of each other and, thus, Fourier signal frequencies were consecutive and could not be resolved. However, when the number of observations was $n = 1000$ and the proportion of smoothness was reduced to $DZ = 0.01$, signal frequencies were within the practical limit for accuracy (0.002) and, thus, the separate frequencies were accurate and clearly resolved. (See Table 3.3.1.)

Kolmogorov-Zurbenko periodograms with Neagu-Zurbenko algorithm smoothing also accurately resolved signal pairs across scenarios with $n = 5000$, regardless of whether the proportion of smoothness was *NZ* = 0.05 or *NZ* = 0.01. In these scenarios, top observed frequencies were within the practical limit for accuracy of $0.002$ and, thus, the separate frequencies were clearly resolved. With $n = 1000$ and $NZ = 0.05$, the performance of the Neagu-Zurbenko algorithm was poor: frequencies were inaccurate in the first signal pair, frequencies were within 0.002 of each other in the second and fourth signal pairs and could not be resolved, and one frequency was inaccurate in the third signal pair. However, with $n = 1000$ and $NZ = 0.01$, performance improved, with frequencies resolved within the practical limit for accuracy in all signal pairs. These results are on par with the results obtained with DiRienzo-Zurbenko algorithm smoothing. (See Table 3.3.2.)

Thus, it appears Kolmogorov-Zurbenko periodograms with dynamic smoothing using relatively high proportions of smoothing lose their ability to resolve signal frequencies as the number of observations decreases. However, resolution can be regained by using relatively low proportions of smoothing because lower proportions of smoothing initially invoke a narrower smoothing window.

| **RESOLUTION: OBSERVED TOP SIGNAL FREQUENCIES**<br>**Amplitude of Signals = 3.58**<br>**Amplitude of Noise = 16**<br>**Signal-to-Noise Ratio = 0.10** | | | | | | | | | |
|---|---|---|---|---|---|---|---|---|---|
| **Actual Signal Frequencies** | | $n = 5000$ | | | | $n = 1000$ | | | |
| | | ***DZ* = 0.05** | | ***DZ* = 0.01** | | ***DZ* = 0.05** | | ***DZ* = 0.01** | |
| $\lambda_1$ | $\lambda_2$ | $\widehat{\lambda_1}$ | $\widehat{\lambda_2}$ | $\widehat{\lambda_1}$ | $\widehat{\lambda_2}$ | $\widehat{\lambda_1}$ | $\widehat{\lambda_2}$ | $\widehat{\lambda_1}$ | $\widehat{\lambda_2}$ |
| 0.040 | 0.030 | 0.040 | 0.030 | 0.040 | 0.030 | 0.030 | 0.028 | 0.040 | 0.030 |
| 0.040 | 0.032 | 0.040 | 0.032 | 0.040 | 0.032 | 0.036 | 0.038 | 0.040 | 0.032 |
| 0.040 | 0.034 | 0.040 | 0.034 | 0.040 | 0.034 | 0.038 | 0.036 | 0.040 | 0.034 |
| 0.040 | 0.036 | 0.040 | 0.036 | 0.040 | 0.036 | 0.038 | 0.040 | 0.036 | 0.040 |

**Table 3.3.1. Observed Top Frequencies in Kolmogorov-Zurbenko Periodograms using DiRienzo-Zurbenko Algorithm Smoothing across Varying Actual Signal Frequencies, with Signal-to-Noise Ratio = 0.10 and Number of Observations $n = 5000$ or $n = 1000$.**
**NB. Initial Window Width $m_{Initial} = 500$, Number of Iterations $k = 3$, and Proportion of Smoothness $DZ = 0.05$ and $DZ = 0.01$.**

| **RESOLUTION: OBSERVED TOP SIGNAL FREQUENCIES**<br>**Amplitude of Signal = 3.58**<br>**Amplitude of Noise = 16**<br>**Signal-to-Noise Ratio = 0.10** | | | | | | | | | |
|---|---|---|---|---|---|---|---|---|---|
| **Actual Signal Frequencies** | | $\boldsymbol{n = 5000}$ | | | | $\boldsymbol{n = 1000}$ | | | |
| | | ***NZ* = 0.05** | | ***NZ* = 0.01** | | ***NZ* = 0.05** | | ***NZ* = 0.01** | |
| $\boldsymbol{\lambda_1}$ | $\boldsymbol{\lambda_2}$ | $\widehat{\boldsymbol{\lambda_1}}$ | $\widehat{\boldsymbol{\lambda_2}}$ | $\widehat{\boldsymbol{\lambda_1}}$ | $\widehat{\boldsymbol{\lambda_2}}$ | $\widehat{\boldsymbol{\lambda_1}}$ | $\widehat{\boldsymbol{\lambda_2}}$ | $\widehat{\boldsymbol{\lambda_1}}$ | $\widehat{\boldsymbol{\lambda_2}}$ |
| 0.040 | 0.030 | 0.040 | 0.030 | 0.040 | 0.030 | 0.036 | 0.044 | 0.040 | 0.030 |
| 0.400 | 0.032 | 0.040 | 0.032 | 0.040 | 0.032 | 0.038 | 0.040 | 0.040 | 0.032 |
| 0.040 | 0.034 | 0.040 | 0.034 | 0.040 | 0.034 | 0.036 | 0.040 | 0.040 | 0.034 |
| 0.040 | 0.036 | 0.040 | 0.036 | 0.040 | 0.036 | 0.038 | 0.040 | 0.036 | 0.040 |

**Table 3.3.2. Observed Top Frequencies in Kolmogorov-Zurbenko Periodograms using Neagu-Zurbenko Algorithm Smoothing across Varying Signal Frequencies, with Signal-to-Noise Ratio = 0.10 and Number of Observations *N* = 5000 or *N* = 1000.**
**NB. Initial Window Width $\boldsymbol{m_{Initial} = 500}$, Number of Iterations $\boldsymbol{k = 3}$, and Proportion of Smoothness $\boldsymbol{NZ = 0.05}$ and $\boldsymbol{NZ = 0.01}$.**

**3.4 Robustness with Respect to Missing Data**

*Definition and Importance*. The fourth criterion used to assess periodogram performance is robustness with respect to missing data. Here, a periodogram algorithm can be viewed as robust with respect to missingness if it can detect, identify, and separate signal frequencies with a high percentage of missing data. This criterion is important because time series analysts frequently encounter situations where data is missing (e.g., instrument malfunction, human error in data collection).

*Statistical Basis and Limits*. There is no statistical basis to establish a theoretical limit for a viable percent of missing data. However, it is possible to demonstrate periodogram performance with respect to sensitivity, accuracy, and resolution under increasing percentages of missingness.

*Demonstration*. To demonstrate the robustness of the Kolmogorov-Zurbenko periodogram with dynamic smoothing, an analysis was done on a simulated time series dataset in which the signal frequencies were set at $\lambda_1 = 0.040$ and $\lambda_2 = 0.036$, amplitude of noise set at 16, and signal-to-noise ratio set at 0.10, resulting in amplitudes of 3.56 for both signals. As with demonstrations of sensitivity, accuracy, and resolution, two numbers of observations were considered: $n = 5000$ and $n = 1000$; likewise, respective proportions of smoothness were $DZ = 0.05$ and $DZ = 0.01$ or $NZ = 0.05$ and $NZ = 0.01$.

Once a simulated dataset was generated for a given scenario, missingness was introduced utilizing random draws from a binomial distribution (i.e., $0 = not\ missing, 1 = missing$) such that data was missing completely at random and the percentage missing ranged from 0% to 70%, in increments of 10%. Setting window width at $m_{Initial} = 500$ and number of iterations at $k = 3$, in each scenario and at each percentage of missingness it was possible to assess sensitivity at a signal-to-noise ratio of 0.10, accuracy at the practical limit of 0.002, and resolution at the practical limit of 0.004 by determining whether or not the top identified frequencies were the same as the true frequencies of 0.040 and 0.036.

With respect to robustness in the context of missing data, Kolmogorov-Zurbenko periodograms with DiRienzo-Zurbenko smoothing demonstrated respectable performance. When the number of observations was $n = 5000$, both signals were detected (sensitivity) and observed frequencies were exact (accuracy) and separated (resolution) with up to 50% missing for $DZ = 0.05$ and with up to 70% missing for $DZ = 0.01$. When the number of observations was $n = 1000$ and the proportion of smoothness was set at $DZ = 0.05$, performance of the DiRienzo-Zurbenko algorithm was poor: with up to 60% missing, Fourier frequencies were consecutive, so signals could not be resolved, and frequencies were inaccurate with 70% missing. However, when the number of observations remained at $n = 1000$ but the proportion of smoothness reduced to $DZ = 0.01$, signals were again detected (sensitivity) and observed frequencies were again exact (accuracy) and separated (resolution) with up to 50% of the data missing. (See Table 3.4.1.)

Kolmogorov-Zurbenko periodograms with Neagu-Zurbenko smoothing also demonstrated respectable performance. When the number of observations was $n = 5000$, both signals were detected (sensitivity) and observed frequencies were exact (accurate) and separated (resolution) with up to 40% missing for $NZ = 0.05$ and with up to 70% missing for $NZ = 0.01$. When the number of observations was $n = 1000$ and the proportion of smoothness was set at *NZ* = 0.05, performance of the Neagu-Zurbenko algorithm was poor: with up to 60% missing, Fourier frequencies were consecutive and signals could not be resolved and, with 70% missing, frequencies were inaccurate. However, when the number of observations remained at $n = 1000$ but the proportion of smoothness reduced to *NZ* = 0.01, the signal was again detected (sensitivity) and observed frequencies were again exact (accuracy) and separated (resolution) with up to 50% of the data missing. (See Table 3.4.2.)

Given the modest decline in performance observed for proportions of smoothness *DZ* = 0.05 and *NZ* = 0.05 when the number of observations moved from $n = 5000$ to $n = 1000$ and the continued sensitivity, accuracy, and resolution observed for proportions of smoothness *DZ* = 0.01 and *NZ* = 0.01 when the number of observations moved from $n = 5000$ to $n = 1000$, it appears robustness for larger proportions of smoothness degrades as the number of observations decreases. However, this is corrected by using smaller proportions of smoothness. Nevertheless, the overall findings are quite good given the relatively low signal-to-noise ratios used in the respective scenarios.

| **ROBUSTNESS: OBSERVED TOP FREQUENCIES**<br>**Signal Frequencies: $\lambda_1 = 0.040$ and $\lambda_2 = 0.036$**<br>**Amplitude of Signals = 3.58**<br>**Amplitude of Noise = 16**<br>**Signal-to-Noise Ratio = 0.10** | | | | | | | | |
|---|---|---|---|---|---|---|---|---|
| **Percent of Missing Data** | $n = 5000$ | | | | $n = 1000$ | | | |
| | ***DZ* = 0.05** | | ***DZ* = 0.01** | | ***DZ* = 0.05** | | ***DZ* = 0.01** | |
| | $\widehat{\lambda_1}$ | $\widehat{\lambda_2}$ | $\widehat{\lambda_1}$ | $\widehat{\lambda_2}$ | $\widehat{\lambda_1}$ | $\widehat{\lambda_2}$ | $\widehat{\lambda_1}$ | $\widehat{\lambda_2}$ |
| 0 | 0.040 | 0.036 | 0.040 | 0.036 | 0.038 | 0.040 | 0.036 | 0.040 |
| 10 | 0.040 | 0.036 | 0.040 | 0.036 | 0.036 | 0.038 | 0.036 | 0.040 |
| 20 | 0.040 | 0.036 | 0.040 | 0.036 | 0.038 | 0.040 | 0.036 | 0.040 |
| 30 | 0.040 | 0.036 | 0.040 | 0.036 | 0.038 | 0.040 | 0.036 | 0.040 |
| 40 | 0.040 | 0.036 | 0.040 | 0.036 | 0.038 | 0.040 | 0.036 | 0.040 |
| 50 | 0.040 | 0.036 | 0.040 | 0.036 | 0.038 | 0.040 | 0.036 | 0.040 |
| 60 | 0.036 | 0.038 | 0.040 | 0.036 | 0.038 | 0.036 | 0.040 | 0.164 |
| 70 | 0.040 | 0.038 | 0.040 | 0.036 | 0.286 | 0.104 | 0.284 | 0.292 |

**Table 3.4.1. Observed Top Frequencies in the Kolmogorov-Zurbenko Periodogram using DiRienzo-Zurbenko Algorithm Smoothing across Varying Percentages of Missing Data, with Signal-to-Noise Ratio = 0.10 and Number of Observations $n = 5000$ or $n = 1000$. NB. Initial Window Width $m_{Initial} = 500$, Number of Iterations $k = 3$, and Proportion of Smoothness $DZ = 0.05$ and $DZ = 0.01$.**

**ROBUSTNESS: OBSERVED TOP FREQUENCIES**
**Signal Frequencies: $\lambda_1 = 0.040$ and $\lambda_2 = 0.036$**
**Amplitude of Signals = 3.58**
**Amplitude of Noise = 16**
**Signal-to-Noise Ratio = 0.10**

| **Percent of Missing Data** | $n = 5000$ | | | | $n = 1000$ | | | |
|---|---|---|---|---|---|---|---|---|
| | **$NZ$ = 0.05** | | **$NZ$ = 0.01** | | **$NZ$ = 0.05** | | **$NZ$ = 0.01** | |
| | $\widehat{\lambda_1}$ | $\widehat{\lambda_2}$ | $\widehat{\lambda_1}$ | $\widehat{\lambda_2}$ | $\widehat{\lambda_1}$ | $\widehat{\lambda_2}$ | $\widehat{\lambda_1}$ | $\widehat{\lambda_2}$ |
| 0 | 0.040 | 0.036 | 0.040 | 0.036 | 0.038 | 0.040 | 0.036 | 0.040 |
| 10 | 0.040 | 0.036 | 0.040 | 0.036 | 0.038 | 0.040 | 0.036 | 0.040 |
| 20 | 0.040 | 0.036 | 0.040 | 0.036 | 0.040 | 0.038 | 0.036 | 0.040 |
| 30 | 0.040 | 0.036 | 0.040 | 0.036 | 0.040 | 0.038 | 0.036 | 0.040 |
| 40 | 0.040 | 0.036 | 0.040 | 0.036 | 0.038 | 0.040 | 0.036 | 0.040 |
| 50 | 0.040 | 0.034 | 0.040 | 0.036 | 0.040 | 0.038 | 0.036 | 0.040 |
| 60 | 0.038 | 0.040 | 0.040 | 0.036 | 0.038 | 0.036 | 0.040 | 0.164 |
| 70 | 0.038 | 0.036 | 0.040 | 0.036 | 0.286 | 0.288 | 0.284 | 0.292 |

**Table 3.4.2. Observed Top Frequencies in the Kolmogorov-Zurbenko Periodogram using Neagu-Zurbenko Algorithm Smoothing across Varying Percentages of Missing Data, with Signal-to-Noise Ratio = 0.10 and Number of Observations $n = 5000$ or $n = 1000$.**
**NB. Initial Window Width $m_{Initial} = 500$, Number of Iterations $k = 3$, and Proportion of Smoothness $NZ = 0.05$ and $NZ = 0.01$..**

## 4. Simulated Example

Kolmogorov-Zurbenko periodograms with dynamic smoothing used in conjunction with the Kolmogorov-Zurbenko Fourier transform comprise a state-of-the-art method for analyzing time series data. To further demonstrate their sensitivity, accuracy, and resolution as well as predictive power, a simulated time series dataset was constructed with known frequencies and known signal-to-noise ratio. Kolmogorov-Zurbenko periodograms with DiRienzo-Zurbenko algorithm smoothing and with Neagu-Zurbenko algorithm smoothing were used to identify signal frequencies and the Kolmogorov-Zurbenko Fourier transform was used to reconstruct the signal. The predictive power of the reconstructed signal was assessed by computing the proportion of variance in the original signal accounted for by the reconstructed signal; this result, in turn, was compared to the predictive power of an autoregression model, another standard method used by time series analysts. The example concludes with reanalyses of the timeseries dataset with 50% of the data missing, again demonstrating the sensitivity, accuracy, and resolution as well as its predictive power with the reconstructed signal.

### 4.1 Construction of the Time Series Dataset

A time series dataset of variable *X* at time *t* was constructed as:

$$X_t = a_1 \cdot \sin(2\pi\lambda_1 t) + a_2 \cdot \sin(2\pi\lambda_2 t) + N(\mu, \sigma^2) \qquad (4.1.1)$$

setting frequency 1 at $\lambda_1 = 0.084$, with amplitude $a_1 = 1$, frequency 2 at $\lambda_2 = 0.098$, with amplitude $a_2 = 1.5$, and random noise at $N(\mu, \sigma^2) = N(0, 16)$. Thus, the signal-to-noise ratio was 0.203, frequencies were to the thousands place, and frequency separation was 0.014. In generating the dataset, time went from $t = 1$ to $t = 5000$, thus producing 5000 observed values of *X*. An interval of this time series, from $t = 30$ to $t = 80$, is displayed in Figure 4.1.1. Here, one can see its rather chaotic nature, with no apparent pattern visible, because two regular signals close in frequency and in amplitude were combined with a high level of random noise.

### 4.2 Identification of Signals

A standard periodogram algorithm (i.e., R command **periodogram( )** ) is available in the *kzft.R* script file and its mean-centered log was used to identify signals in the time series dataset. The result is displayed in Figure 4.2.1(a). One can see that there are two prominent signal spikes, one at $\lambda_1 = 0.084$ and the other at $\lambda_2 = 0.098$. However, there are a multitude of minor spikes and it is not clear whether these are actual signals or random fluctuations in the log-periodogram. To gain a clearer notion of what signals are actually present, two Kolmogorov-Zurbenko periodograms were constructed, one using DiRienzo-Zurbenko algorithm smoothing and one using Neagu-Zurbenko algorithm smoothing, both with $Proportion\ of\ Smoothing = 0.01$, $m_{Initial} = 500$, and $k = 3$, and are displayed in Figure 4.2.1(b) and Figure 4.2.1(c), respectively. In both Kolmogorov-Zurbenko periodograms, one can see that the multitude of minor spikes have dissolved and those remaining are muted.

While the standard log-periodogram is ambiguous with respect to the presence of weaker signals, the Kolmogorov-Zurbenko periodograms with dynamic smoothing demonstrate sensitivity, accuracy, and resolution, both clearly identifying $\lambda_1 = 0.084$ and $\lambda_2 = 0.098$ with less ambiguity. Given both Kolmogorov-Zurbenko periodograms identified the same signal frequencies, it appeared possible to reconstruct of the original signal.

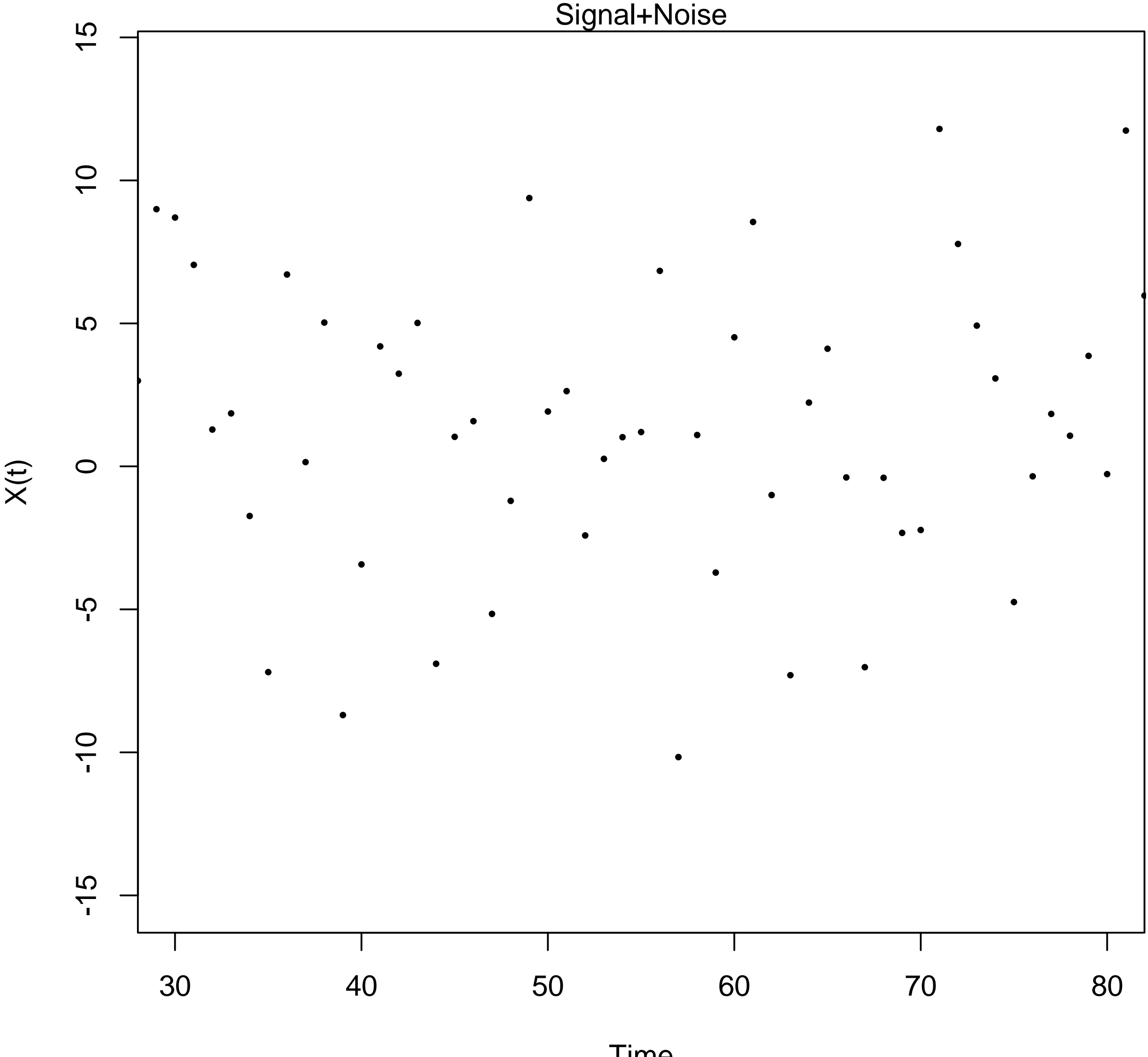

**Figure 4.1.1. Time Series Plot between *t*=30 and *t*=80 for Complete Data where:**
$\boldsymbol{X(t) = 1 \cdot \sin(2\pi(0.084)t) + 1.5 \cdot \sin(2\pi(0.098)t) + N(0,16)}$.

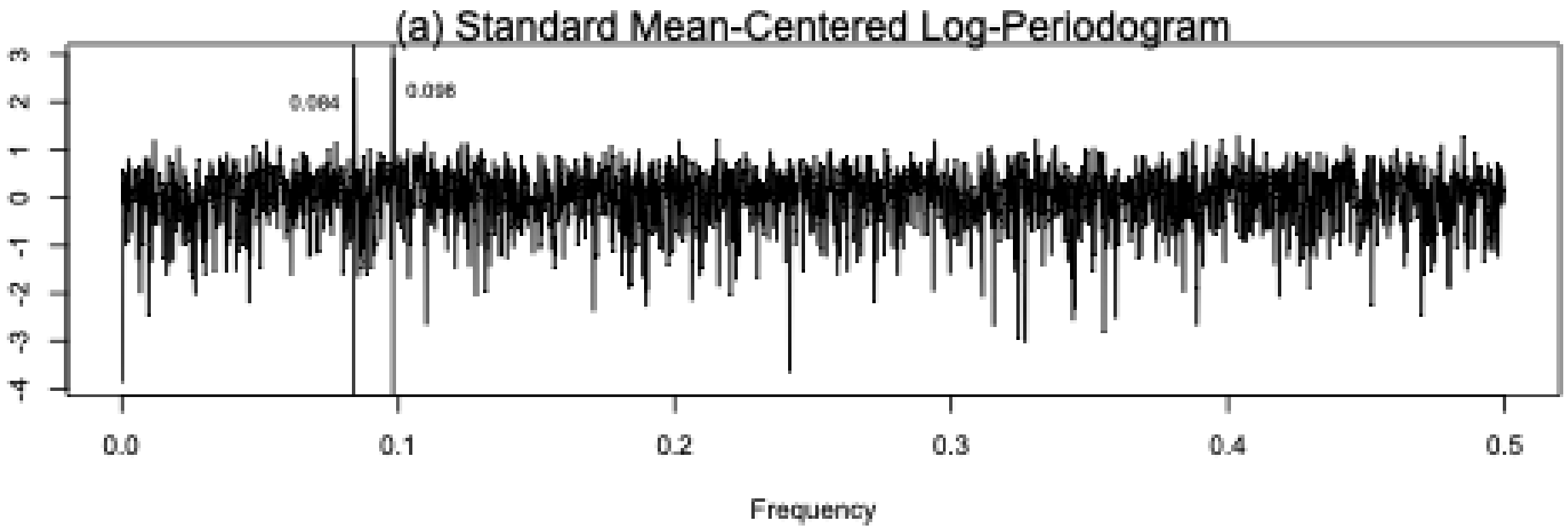

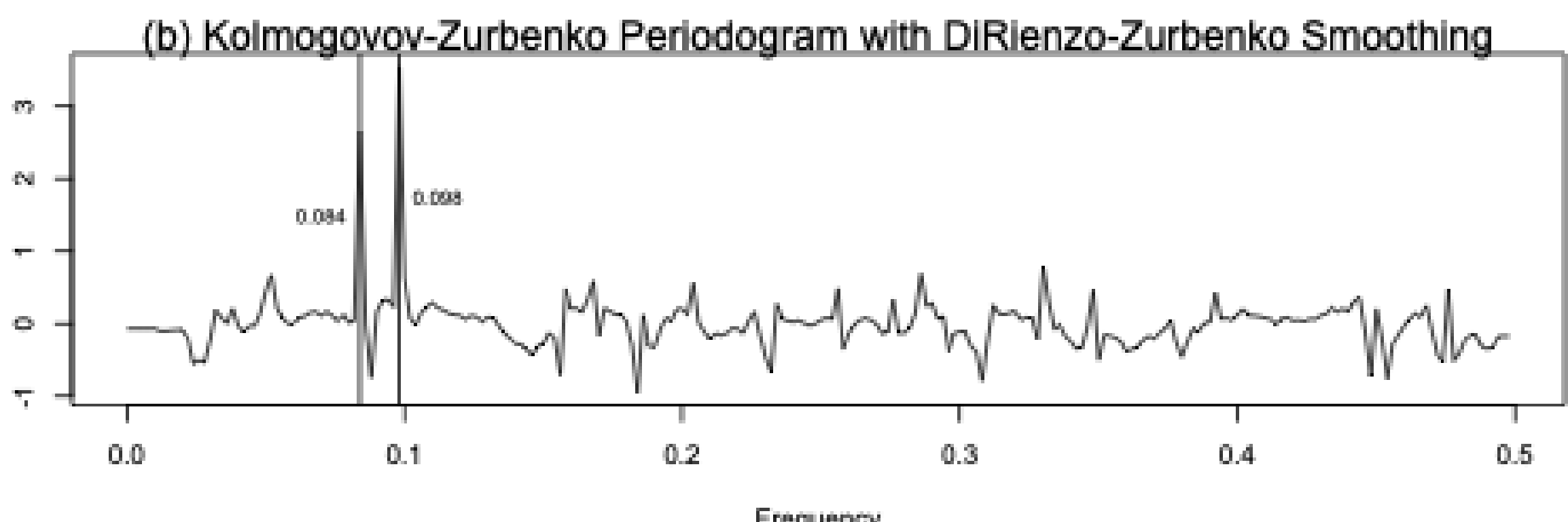

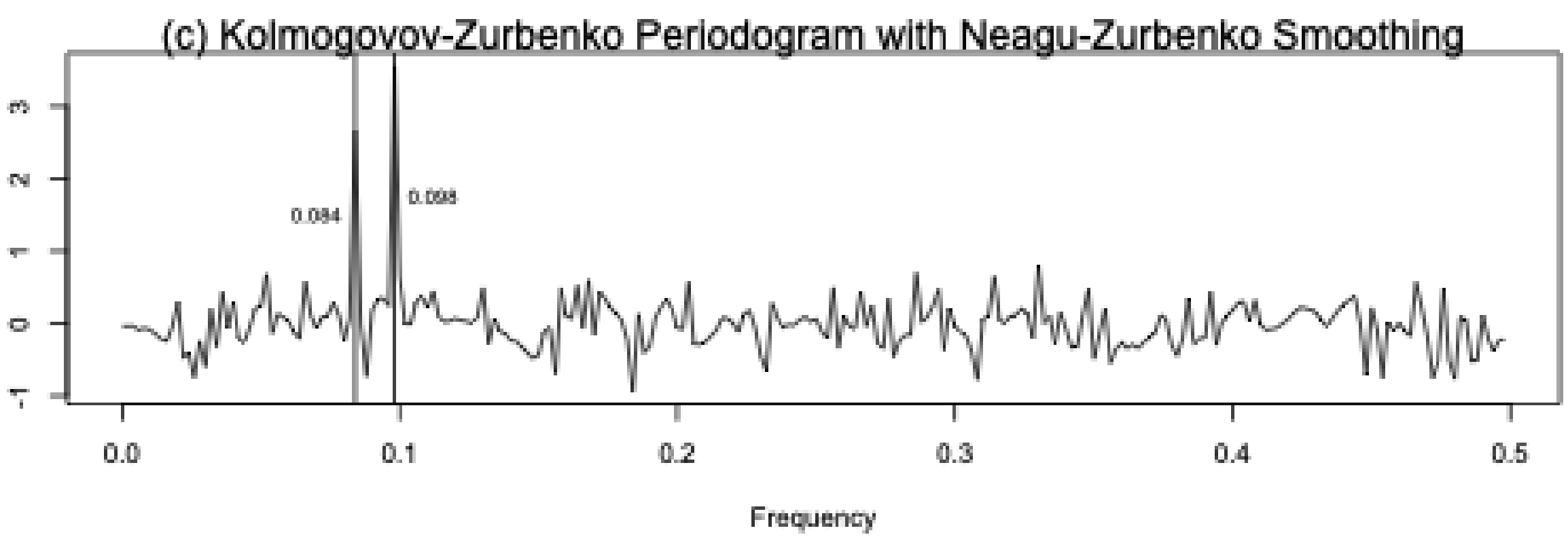

**Figure 4.2.1. Spectral Analyses using Complete Data**

**(a) Standard Mean-Centered Log-Periodogram**

**(b) Kolmogorov-Zurbenko Periodogram with DiRienzo-Zurbenko Algorithm Smoothing**

**(c) Kolmogorov-Zurbenko Periodogram with Neagu-Zurbenko Algorithm Smoothing**

**4.3 Reconstruction of Original Signal**

In reconstructing the original time series signal, the goal is to correctly predict the values of variable *X* over time, using frequencies identified by the Kolmogorov-Zurbenko periodograms with dynamic smoothing. To reconstruct the signal, the Kolmogorov-Zurbenko Fourier transform was used to construct two signals, one with $\lambda_1 = 0.084$ and the other with $\lambda_2 = 0.098$. These two signals were added together, then the real-valued component extracted and multiplied by 2 to form the reconstructed signal. It must be remembered that the Kolmogorov-Zurbenko Fourier transform generates complex numbers; thus, the real part must be extracted and multiplied by 2 to obtain the estimated value of *X* for each value of *t*.

For the time series interval running from $t = 30$ to $t = 80$, Figure 4.3.1 displays the original data (i.e., signal plus noise), the original signal (i.e., signal only), and the reconstructed signal. Despite the fact that the original signal cannot be easily identified in the plot of the original data (signal plus noise), it can be seen that the reconstructed signal runs a very close approximation to the original signal (signal only). Further, the predictive power of the reconstructed signal is strong. Across the interval from $t = 4$ to $t = 5000$, the correlation between the original signal and the reconstructed signal is $r = 0.977$, thus accounting for 95.4% of the variance in the original signal. It should be noted that the first three observations are lost due to averaging within the algorithm.

**4.4 Comparison to Autoregression**

Autoregression models are used extensively in time series analysis. However, the approach using the Kolmogorov-Zurbenko periodogram with dynamic smoothing followed with signal reconstruction using the Kolmogorov-Zurbenko Fourier transform is far superior to autoregression modeling in relation to prediction with a sinusoidal time series dataset. To demonstrate this, autoregression analysis was done on the time series dataset. The standard autoregression algorithm in R is **ar( )** and can be found in the *stats* package. It identified an autoregression model of order 12, meaning that the values of *X* at time *t* correlated with subsequent values of *X* up to 12-time units ahead. However, correlations were extremely weak, with magnitudes ranging from *0.021* at *t+8* to *0.125* at *t+6*. Figure 4.4.1 presents the related correlogram.

The autoregression analysis indicated that the variance that could not be accounted for in the model was *16.578*, while the total variance of the time series dataset was *17.403*. As a result, the autoregression analysis could not account for 95.3% of the total variance. Thus, the autoregression model accounted for only 4.7% of the *total variance in the time series dataset*, while the reconstructed signal based on approaches using the Kolmogorov-Zurbenko periodogram with dynamic smoothing in conjunction with the Kolmogorov-Zurbenko Fourier transform accounted for 95.4% of the variance in the *actual signal*. The autoregression model for the signal provided above resembles mostly heavy noise with no practical opportunity to predict the periodic signal buried within that noise. This situation is common in practice and can be resolved by utilizing the Kolmogorov-Zurbenko periodogram with dynamic smoothing to identify hidden signal frequencies, followed by use of the Kolmogorov-Zurbenko Fourier transform to reconstruct the hidden signal, as was demonstrated above.

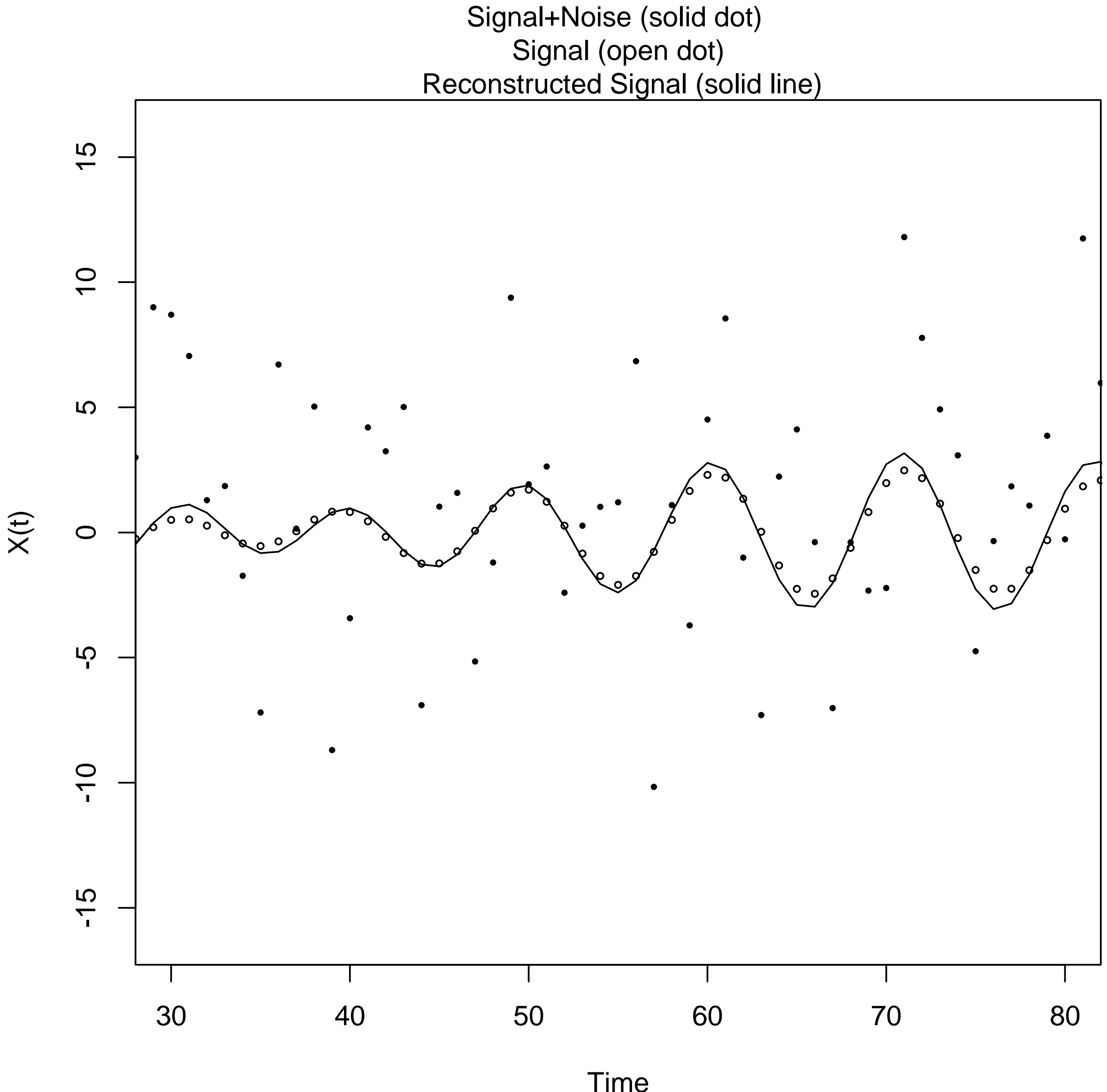

**Figure 4.3.1. Plots of signal+noise, signal, and reconstructed signal using the Kolmogorov-Zurbenko Fourier transform (*m*=500 and *k*=3).**

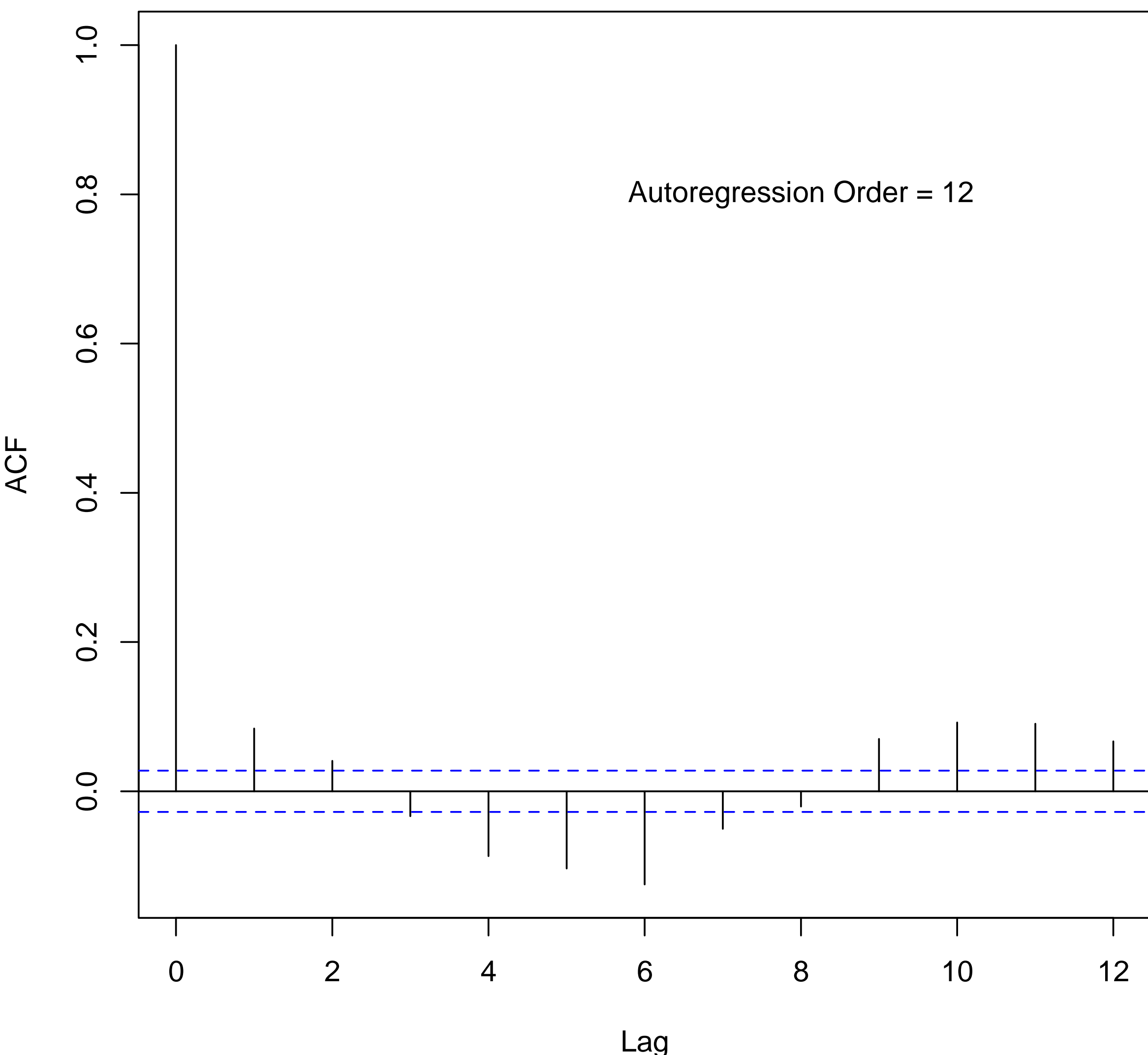

**Figure 4.4.1. Correlogram of Autoregression Analysis for Simulated Time Series for Complete Data**

**4.5 Robustness with Respect to Missing Data**

Identification of signal frequencies and reconstruction of the original signal was repeated, but with 50% of the datapoints missing. As in the earlier demonstration of robustness in the context of missing data, this began with the complete time series dataset and missingness was introduced using random draws from a binomial distribution (i.e., $0 = not\ missing$, $1 = missing$) so that data were missing completely at random with the percentage set at 50%. Next, two Kolmogorov-Zurbenko periodograms, one with DiRienzo-Zurbenko algorithm smoothing and one with Neagu-Zurbenko algorithm smoothing, were constructed and are displayed in Figure 4.5.1(a) and Figure 4.5.1(b), respectively. Despite a missingness rate of 50%, sensitivity, accuracy, and resolution were maintained for both periodograms and, again, two signal spikes were evident, one at $\lambda_1 = 0.084$ and the other at $\lambda_2 = 0.098$.

To reconstruct the original signal, the Kolmogorov-Zurbenko Fourier transform was again used to construct two signals, one at $\lambda_1 = 0.084$ and the other at $\lambda_2 = 0.098$, which were, in turn, added together with the real-valued component extracted and multiplied by 2. Figure 4.5.1(c) displays the original data (signal plus noise), original signal (signal only), and the reconstructed signal, with the former two plots notably missing values of $X_t$. Again, one can see that the reconstructed signal closely approximates the original signal.

The predictive power of the reconstructed signal was also strong. Across the available data, the correlation of the reconstructed signal with the original data was 0.954, thus accounting for 91.0% of the variance in the original signal.

**4.6 Summary of Simulated Example**

This section demonstrated the ease of use and predictive power of an approach to time series analysis that utilizes Kolmogorov-Zurbenko periodograms with dynamic smoothing to detect, identify, and separate signal frequencies and the Kolmogorov-Zurbenko Fourier transform to reconstruct the original time series. The utility and power of this approach is apparent, even when signals are weak relative to random noise (i.e., low signal-to-noise ratio), even when signals are close in frequency and amplitude, and even when there is a substantial amount of missing data.

When periodicity is present in a time series dataset, the predictive power of this approach is quite strong compared to another standard method of time series analysis, the autoregression model. Not only are these spectral analytic approaches good in predicting values of an underlying set of signals, but they are also able to do so into perpetuity, assuming no catastrophic event impacts the time series process.

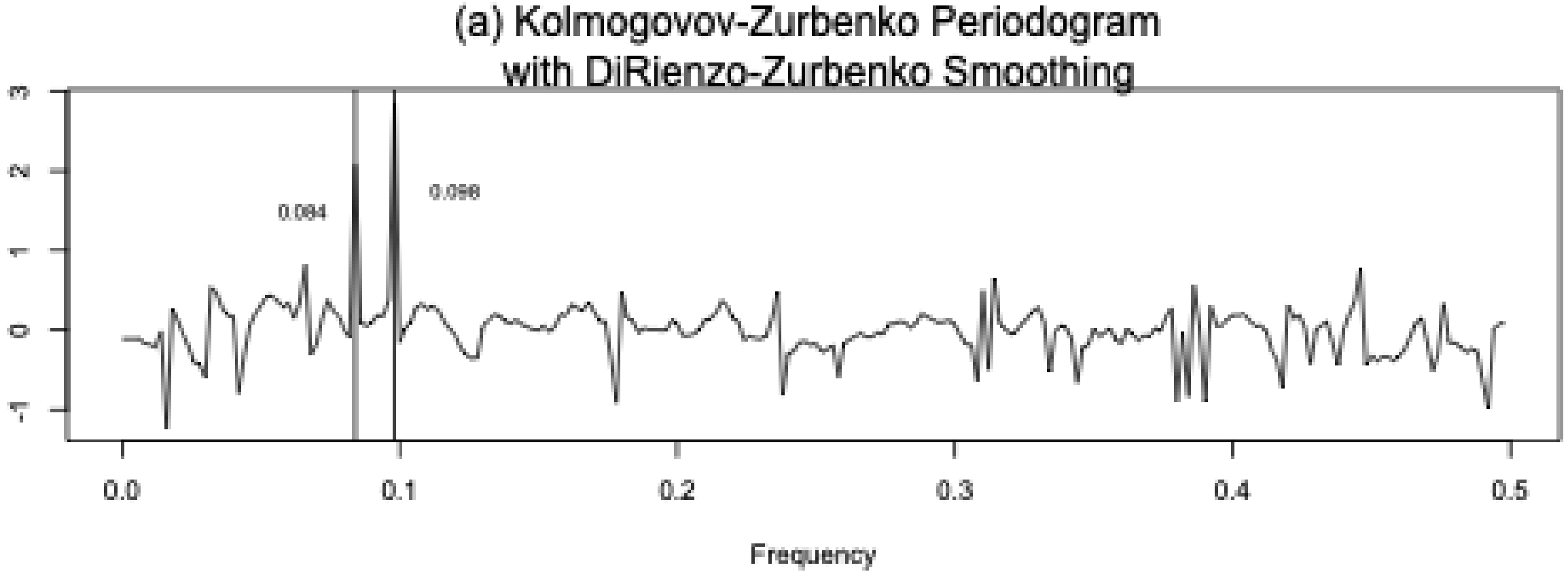

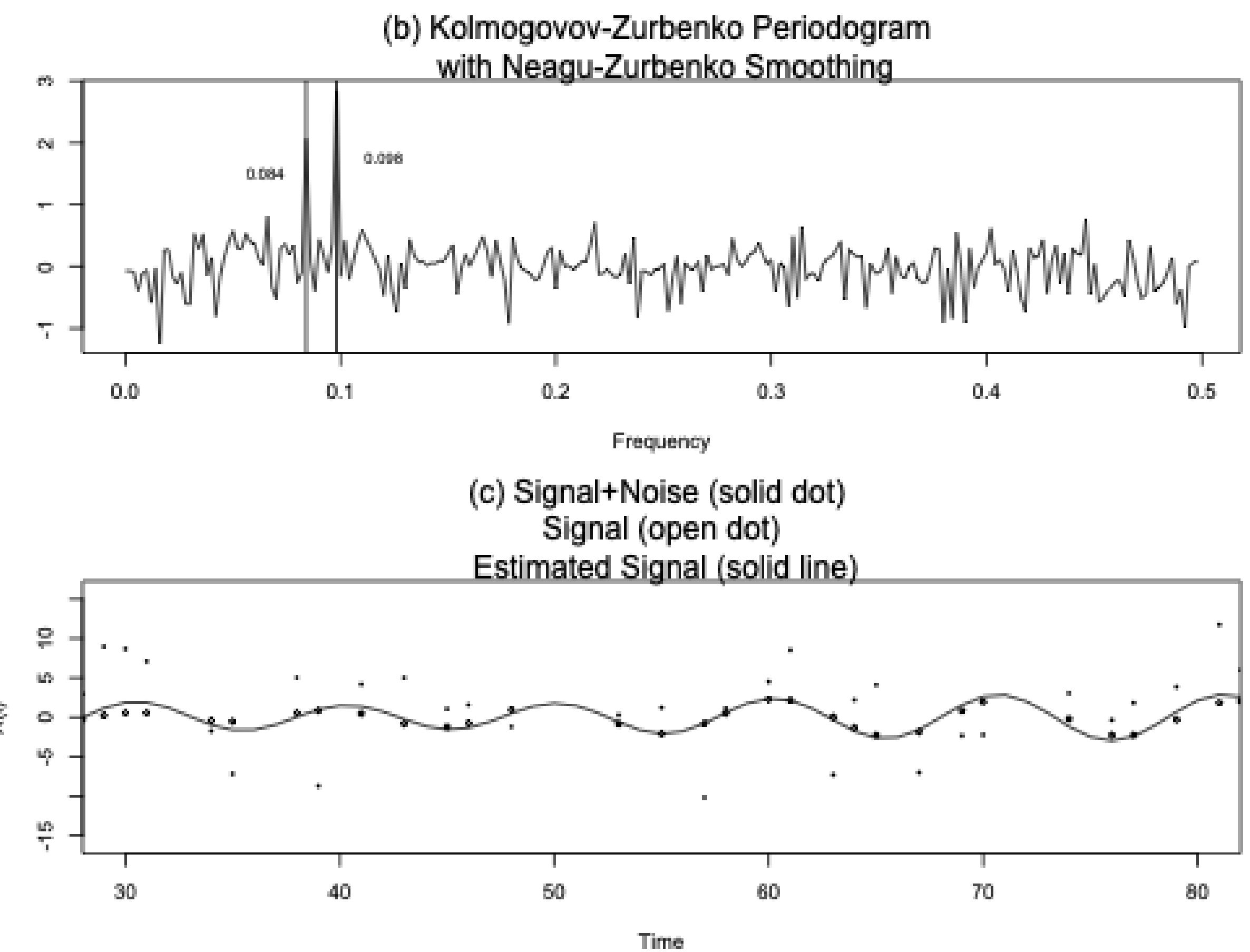

**Figure 4.5.1. Time Series Analysis and Signal Reconstruction with 50% Missing Data.**

- **(a) Kolmogorov-Zurbenko periodogram using DiRienzo-Zurbenko algorithm smoothing ($m_{Initial} = 500$, $k = 3$, and $DZ = 0.01$)**
- **(b) Kolmogorov-Zurbenko periodogram using Neagu-Zurbenko algorithm smoothing ($m_{Initial} = 500$, $k = 3$, and $NZ = 0.01$)**
- (c) **Plots of signal+noise, signal, and estimated signal using the Kolmogorov-Zurbenko Fourier transform ($m = 500$ and $k = 3$)**

## 5. Examples in Current Research

There are a range of studies, in fields as diverse as earth and atmospheric sciences as well as public health, that have utilized Kolmogorov-Zurbenko periodograms with DiRienzo-Zurbenko algorithm smoothing to identify heretofore hidden periodicities in time series data. For example just as oceans have oceanic tides, the atmosphere has its own tides and Zurbenko and Potrzeba (2010, 2013) have used this approach to identify important patterns in atmospheric tides and atmospheric tidal waves, highlighting critical patterns (signals) that can attenuate or can exacerbate the destructive power of hurricanes. In regard to climate, these same researchers utilized the Kolmogorov-Zurbenko periodogram with DiRienzo-Zurbenko algorithm smoothing to identify patterns related to climate change: they found that cycles in the planetary orbits have an impact on sunspot activity (Potrzeba-Macrina & Zurbenko, 2019) and, in turn, were able to assess the relative impact of sunspot activity, solar irradiation, and human factors on temperature (Potrzeba-Macrina & Zurbenko, 2020; Zurbenko & Potrzeba-Macrina, 2019b, 2019a). In the field of public health, both Close and Zurbenko (2008) as well as Zurbenko and Pera (2020) utilized Kolmogorov-Zurbenko periodograms with DiRienzo-Zurbenko algorithm smoothing to identify seasonal, weekly, and daily patterns in traffic fatalities, with the latter pair of researchers finding increased fatalities related to use of daylight savings time. The reader is encouraged to review these original works to gain a better understanding of how this approach has been used as well as the power of the methods themselves.

## 6. Conclusions

This article began with summaries of the statistical foundations for Kolmogorov-Zurbenko periodograms with DiRienzo-Zurbenko algorithm smoothing and with Neagu-Zurbenko algorithm smoothing. Because the logarithmic transformation of a periodogram offers distinct advantages over a standard periodogram, Kolmogorov-Zurbenko periodograms are, by default, log-periodograms. Typical methods used to construct smoothed periodograms utilize a static spectral window with fixed width, but both the DiRienzo-Zurbenko algorithm and the Neagu-Zurbenko algorithm use a dynamic spectral window, zooming in (decreasing the window width) when a signal frequency is present and zooming out (increasing the window width) when signal frequencies are not present. As a result, Kolmogorov-Zurbenko periodograms with dynamic smoothing more clearly identify hidden signals and represent significant improvements in the construction of periodograms.

While the DiRienzo-Zurbenko algorithm adapts window size according to the proportion of total variance, the Neagu-Zurbenko algorithm adapts window size according to the proportion of total departure from linearity. Instructions for accessing the Kolmogorov-Zurbenko periodogram function within the statistical platform $R$, along with basic instructions in its use, were provided. It was noted that spectral analysts who prefer working in terms of proportion of total variance can invoke the DiRienzo-Zurbenko algorithm and those who prefer to work in terms of proportion of total departure from linearity can choose the Neagu-Zurbenko algorithm.

The theoretical and practical limits of the Kolmogorov-Zurbenko periodogram with dynamic smoothing were established for sensitivity (ability to detect weak signals), accuracy (ability to correctly identify signal frequencies), resolution (ability to resolve signals close in frequency), and robustness (sensitivity, accuracy, and resolution when data is missing) and each included demonstrations using simulated datasets. For a given number of observations, $n$, signal-to-noise ratio, $s^2/n^2$, initial window width, $m_{Initial}$, and proportion of smoothness, $DZ$ or $NZ$, the theoretical limit of sensitivity for DiRienzo-Zurbenko algorithm is $s^2/n^2 \geq DZ$ and the practical limit of signal sensitivity for the Neagu-Zurbenko algorithm is some $s^2/n^2 < NZ$. For both the DiRienzo-Zurbenko algorithm and the Neagu-Zurbenko algorithm, their theoretical limit for frequency accuracy is $\pi/n$, with a practical limit of $\pi/m_{Initial}$, and their theoretical limit for frequency resolution is $2\pi/n$, with a practical limit of $2\pi/m_{Initial}$. In establishing these theoretical limits, it was shown that Kolmogorov-Zurbenko periodograms with dynamic smoothing exceed the Rose criteria for sensitivity and have accuracy and resolution as good as or better than typical log-periodograms that use static smoothing

With respect to the practical limits for robustness, Kolmogorov-Zurbenko periodograms with dynamic smoothing demonstrated sensitivity, accuracy, and resolution with up to 50% of the data missing for the DiRienzo-Zurbenko algorithm and with up to 40% of data missing for the Neagu-Zurbenko algorithm, except when the number of observations was relatively small ($n = 1000$) and the proportion of smoothness was relatively high ($DZ = 0.05$ or $NZ = 0.05$).

The demonstrated performance of both the DiRienzo-Zurbenko algorithm smoothing and Neagu-Zurbenko algorithm smoothing was quite good across all criteria. With regard to sensitivity, the Neagu-Zurbenko algorithm performed even better than the DiRienzo-Zurbenko algorithm when the number of observations was relatively small and the proportion of smoothness was relatively low. With regard to accuracy, performance of the DiRienzo-Zurbenko algorithm was on par with the Neagu-Zurbenko algorithm, although when the number of observations was small and the proportion of smoothness high, the DiRienzo-Zurbenko algorithm performed better. With regard to resolution, the DiRienzo-Zurbenko algorithm and the Neagu-Zurbenko algorithm both performed well, except when the number of observations was relatively small and the proportion of smoothness was relatively high. With regard to robustness in the face of missing data, neither the DiRienzo-Zurbenko algorithm nor the Neagu-Zurbenko algorithm performed well when number of observations was small and proportion of smoothness was high; in all other scenarios, both algorithms maintained sensitivity, accuracy, and resolution, with up to 50% of data missing for the DiRienzo-Zurbenko algorithm and up to 40% of data missing for the Neagu-Zurbenko algorithm.

It is worth noting that the DiRienzo-Zurbenko algorithm and the Neagu-Zurbenko algorithm evidenced their poorest performance across the criteria when the number of observations was relatively small ($n = 1000$ versus $n = 5000$) and proportion of smoothness was relatively high ( ($DZ$ or $NZ = 0.05$ versus $DZ$ or $NZ = 0.01$). For both algorithms, this implies that when the number of observations is relatively large, the analyst can invoke a higher proportion of smoothness. However, when the number of observations is relatively small, the analyst should use a lower proportion of smoothness so as to better detect, identify, and resolve signal frequencies in the context of high amounts of noise because the lower proportion of smoothing results in a narrower smoothing window width in the presence of a signal.

An example was also presented using a simulated time series dataset in which two signals close in frequency and amplitude were embedded in a significant level of random noise. Kolmogorov-Zurbenko periodograms, one using DiRienzo-Zurbenko algorithm smoothing and one using Neagu-Zurbenko algorithm smoothing, accurately identified the same two signal frequencies and the Kolmogorov-Zurbenko Fourier transform recreated the original combined signal. The reconstructed signal had a high level of predictive power, accounting for 95.4% of the variance in the original signal using the complete dataset and accounting for 91.0% of the variance in the original signal when 50% of the data were missing. By comparison, an autoregression model of order 12 only accounted for 4.7% of the variance present in the original dataset (signal-plus-noise). Thus, an approach using the Kolmogorov-Zurbenko periodogram with dynamic smoothing to identify hidden signal frequencies followed by signal reconstruction using the Kolmogorov-Zurbenko Fourier transform represents a state-of-the-art approach that is straightforward in its use and produces results that have a considerably high level of predictive power when periodicity is present in a time series dataset.

Although periodogram sensitivity, accuracy, and resolution are necessary conditions for detecting, identifying, and separating signal frequencies, they are not sufficient for framing a model to predict future values of a variable. In addition, a periodogram must be able to estimate signal strength with adequate precision. This issue is important because both signal frequency and signal strength are needed to provide a comprehensive estimate for future values of a variable in a periodic time series. As a result, we need a way to compare signal strength estimate precision of Kolmogorov-Zurbenko periodograms utilizing dynamic smoothing (i.e., DiRienzo-Zurbenko algorithm, Neagu-Zurbenko algorithm) to such estimate precision of log-periodograms utilizing static smoothing windows. To this end, our next investigation will make this comparison by contrasting the widths of their respective signal strength confidence intervals; here, the notion is that the narrower the confidence interval is when a signal is present, the more precise the estimate of signal strength. Consequently, our next article will provide an overview of the derivation of confidence intervals for signal strength, present an algorithm to compute them when dynamic smoothing is used, and establish, compare, and contrast the theoretical and practical limits of confidence interval widths for Kolmogorov-Zurbenko periodograms with dynamic smoothing to log-periodograms with static smoothing.